\documentclass[reprint,superscriptaddress,amsmath,amssymb,twocolumn]{revtex4-2}

\usepackage{amsmath}
\usepackage{mathtools}
\usepackage{amssymb}
\usepackage{graphicx}
\usepackage{bm}
\usepackage[colorlinks, linkcolor=blue, citecolor=blue, urlcolor=blue, breaklinks=true]{hyperref}
\usepackage{color,dsfont,multirow,booktabs}
\usepackage{braket}
\usepackage{tabularx}
\usepackage{lipsum,booktabs}
\usepackage{subfiles}
\usepackage{subcaption}
\usepackage{caption}
\usepackage[dvipsnames]{xcolor}
\usepackage{ragged2e}
\usepackage{orcidlink}
\usepackage[T1]{fontenc}
\usepackage[normalem]{ulem}

\newcommand{\commentold}[1]{}
\DeclareMathSymbol{:}{\mathpunct}{operators}{"3A}

\newcommand\numberthis{\addtocounter{equation}{1}\tag{\theequation}}
\newcommand{\figpanel}[2]{\hyperref[#1]{\ref*{#1}(#2)}}

\begin{document}
\date{\today}

\title{Reservoir-Engineered Low-Threshold Quantum Energy Storage}

\author{Borhan Ahmadi\orcidlink{0000-0002-2787-9321}}
\email{borhan.ahmadi@ug.edu.pl}
\address{International Centre for Theory of Quantum Technologies, University of Gdańsk, ul. prof. Marii Janion 4, 80-309 Gdańsk, Poland}
\author{André H. A. Malavazi\orcidlink{0000-0002-0280-0621}}
\email{andrehamalavazi@gmail.com}
\address{International Centre for Theory of Quantum Technologies, University of Gdańsk, ul. prof. Marii Janion 4, 80-309 Gdańsk, Poland}
\author{Paweł Mazurek\orcidlink{0000-0003-4251-3253}}
\address{Institute of Informatics, Faculty of Mathematics, Physics and Informatics, University of Gdańsk, Wita Stwosza 63, 80-308 Gdańsk, Poland}
\author{Shabir Barzanjeh}
\address{Department of Physics and Astronomy, University of Calgary, Calgary, AB T2N 1N4 Canada}
\author{Paweł Horodecki}
\address{International Centre for Theory of Quantum Technologies, University of Gdańsk, ul. prof. Marii Janion 4, 80-309 Gdańsk, Poland}

\begin{abstract}
Fast charging of quantum batteries requires a mechanism that amplifies the energy transferred to the storage mode without relying on uncontrolled gain or phenomenological non-Hermitian dynamics. Inspired by the broken/unbroken structure of parity--time-symmetric systems, we introduce a reservoir-engineered quantum battery in which a charger and a battery mode are coupled through a dissipative mediator mode and driven by a two-photon pump. Eliminating the fast mediator yields a reduced two-mode Lindblad model with a complex dissipative coupling and renormalized damping rates. The resulting drift matrix has a pump-induced stability threshold: below threshold, the seeded response remains bounded, whereas above threshold, a weak seed excites a growing mode and the battery occupation increases exponentially. We show that this threshold is substantially lower than in a coherent beam-splitter charger--battery benchmark, implying a quadratic reduction of the required critical pump power. For the representative parameters studied here, the dissipative architecture reaches the broken regime with about \(61\%\) less critical pump power, opening a pump-power window in which dissipative charging is exponential while the coherent benchmark remains below pump threshold. Crucially, in the broken dissipative regime the growth is dominated by the seed-selected coherent battery displacement, rather than by incoherent fluctuation buildup, making a large fraction of the stored energy directly extractable by a displacement operation. The broken-regime boundary is a dynamical stability threshold, not generally an exceptional point, and the full three-mode Lindblad model confirms the validity of the reduced two-mode description in the controlled fast-mediator regime. Our results provide a completely positive, trace-preserving route to pump-efficient, low-threshold, and coherently addressable quantum energy storage using engineered reservoirs.
\end{abstract}

\maketitle
\noindent 
\textit{Introduction}--
Open quantum systems can display dynamical transitions that have no direct counterpart in closed Hermitian evolution. A prominent example is provided by parity--time (PT) symmetric systems, where the interplay of gain and loss separates the spectrum into two regimes: an unbroken regime, in which the relevant modes remain dynamically stable, and a broken regime, in which one or more modes acquire exponential growth or decay \cite{Ruter2010PTSymmetry,ElGanainy2018NonHermitian,Ozdemir2019EPReview,Ashida2020NonHermitianReview}. This organizing principle has led to striking effects in optics, photonics, and condensed-matter platforms, including unidirectional invisibility \cite{Feng2013Unidirectional}, mode-selective lasing \cite{Feng2014SingleModeLaser,Hodaei2014SingleModeLaser}, coherent perfect absorption \cite{PhysRevLett2019,changqing2021science}, and enhanced control of energy-transfer pathways in cavity and circuit QED \cite{Naghiloo2019QuantumEP,PhysRevA.108.022215}. In ideal PT-symmetric models, the boundary between the two regimes often occurs at an exceptional point, where both eigenvalues and eigenvectors coalesce. More generally, however, driven-dissipative quantum systems are governed by a drift matrix whose stability is determined by its spectral abscissa. In this broader Lyapunov sense, the transition to a broken regime occurs when the largest real part of the drift eigenvalues changes sign, so that the zero-displacement solution loses linear stability \cite{Lyapunov1992Stability,Khalil2002NonlinearSystems,Trefethen2005Spectra}. This stability threshold need not coincide with an exceptional point or with a strict PT-symmetry-breaking transition. This distinction is central to the mechanism developed below.

Quantum batteries (QBs) \cite{PhysRevE.87.042123,ferraro2026opportunities,camposeo2025quantum,RevModPhys.96.031001,PhysRevResearch.2.013095,bv4w-jr6q,ahmadi2025harnessing,eabk3160,PhysRevApplied.14.024092} are quantum systems designed to store and release energy using coherent, collective, or correlation-assisted processes \cite{PhysRevE.102.052109,PhysRevLett.131.030402,PhysRevLett.120.117702,PhysRevA.107.042419,6kwv-z6fx,PhysRevResearch.5.013155,Rodriguez_2024,8xsm-5mb6,PhysRevA.109.042207,PhysRevLett.134.180401,zhang2024quantum,PhysRevLett.122.210601,PhysRevA.102.052223,PhysRevA.104.032207,PhysRevA.105.062203,PhysRevE.104.064143,PhysRevLett.132.210402,PhysRevApplied.23.024010,Shastri2025,PhysRevA.111.042216,malavazi2025charge,zakavati2025optimizing,ahmadi2026ChiralSqueezing}. A central goal is to charge them rapidly while retaining a useful, controllable form of stored energy. PT-inspired and non-Hermitian approaches suggest one possible route: if the battery dynamics can be brought into a broken regime, the response to a weak seed can grow exponentially, leading to rapid energy accumulation. Existing proposals, however, often rely on effective non-Hermitian Hamiltonians, explicit gain media, or finely balanced gain--loss dimers \cite{8xsm-5mb6,PhysRevA.109.042207}. Such ingredients may obscure the microscopic origin of the supplied energy, amplify noise, and require delicate calibration in realistic quantum devices.

Here we propose a fully open-system route to low-threshold quantum energy storage based on reservoir-engineered dissipative interactions. Reservoir engineering turns the environment from an uncontrolled source of decoherence into a designed dynamical resource: by tailoring the system--bath couplings, one can make dissipation generate selected interactions, stabilize desired states, or amplify selected dynamical modes \cite{Poyatos1996ReservoirEngineering,Verstraete2009Dissipation,PhysRevLett.112.133904,PhysRevX.5.021025,PRXQuantum.4.010306}. The specific resource exploited below is dissipative interference, the same broad mechanism underlying quantum-noise cancellation in dissipative optomechanics, level attraction in magnon--photon systems, and linewidth control through lossy ancillary modes \cite{PhysRevLett.102.207209,PhysRevLett.121.137203,PhysRevLett.126.163604}. In our setting, the charger and the battery are not coupled by a direct beam-splitter Hamiltonian. Instead, a two-photon-driven charger mode is connected to the battery through a fast lossy mediator and two engineered nonlocal reservoir channels. Eliminating the mediator gives a reduced two-mode Lindblad model in which the charger--battery coupling appears entirely in the dissipative kernel. The same elimination also renormalizes the effective damping rates. Physically, this renormalization originates from dissipative interference: the direct loss channel and the indirect pathway through the mediator contribute coherently at the level of the Kossakowski matrix, producing both a complex off-diagonal dissipative coupling and a negative correction to the diagonal decay rates. The reduced dynamics remains completely positive and trace preserving, but its pump-driven drift matrix can undergo a Lyapunov instability at a much lower pump amplitude than a coherent exchange benchmark. Thus the broken/unbroken terminology used below refers to dynamical stability of the first-moment drift, not to a strict PT-symmetry-breaking transition.

The main result is that the reduced dissipative architecture enters the broken regime at a much smaller pump amplitude than an equivalent coherent beam-splitter charger--battery benchmark. This lower threshold originates from dissipative interference: the eliminated mediator contributes both an effective complex coupling between charger and battery and a negative correction to the effective damping rates. As a result, the pump strength required for the largest Lyapunov exponent \(\max_i{\Re}[\lambda_i(\mathcal A(\mathcal E))]\) of the drift matrix $\mathcal A(\mathcal E)$ associated with the model to become positive is strongly reduced. The broken regime should therefore be understood as a pump-induced dynamical instability of the physical drift matrix, not as an exceptional point of the passive spectrum~\cite{PhysRevA.109.042207,8xsm-5mb6,ll76-j2l5}. In the parameter regimes studied below, the exceptional-point condition and the broken-regime threshold are distinct.

This distinction leads to a practical advantage. Below threshold, the seeded response remains bounded, and the battery occupation saturates. Above threshold, the same weak seed excites a growing mode, and the battery occupation increases exponentially. Compared with coherent exchange, the dissipative model reaches this regime for substantially weaker pumping and produces a cleaner charging response with reduced reversible oscillations. The full three-mode dissipative model is used as a microscopic consistency check: in the fast-mediator regime it reproduces the reduced dynamics, confirming that the low-threshold enhancement is not an artifact of adiabatic elimination. The required ingredients are compatible with opto-electromechanical \cite{Toth2017,Barzanjeh2017}, superconducting \cite{PhysRevApplied.4.034002,PhysRevX.7.041043}, and magnonic \cite{PhysRevLett.123.127202,2303.04358} platforms, where engineered reservoirs, lossy auxiliary modes, and dissipative couplings are experimentally available. Recent dissipative-coupling phonon and magnon-polariton lasing experiments further show that such loss-mediated interactions can be used as threshold-control resources rather than as mere decoherence channels \cite{Zhang2022PhononLasing,Wang2025MagnonPolaritonLasing}.
\begin{figure}[t]
\center
\includegraphics[width=1\columnwidth]{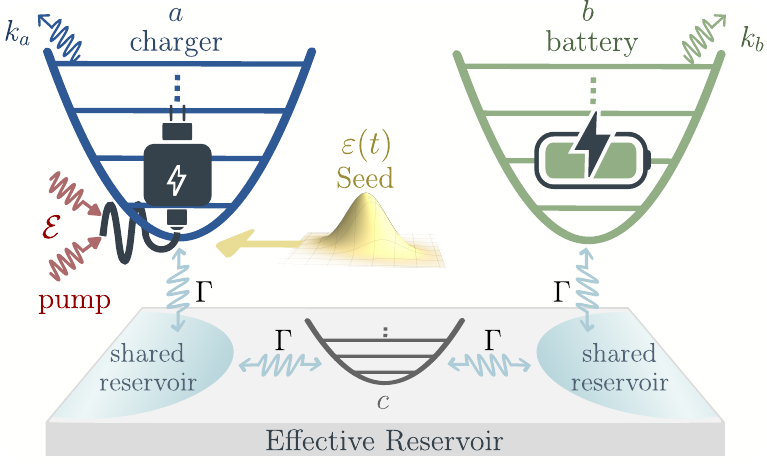}
    \caption{Schematic of the dissipative quantum battery.
    The charger mode \(a\), with resonance frequency \(\omega_a\) and damping rate \(\kappa_a\), is driven by a two-photon pump of physical frequency \(2\omega_L\) and amplitude \(\mathcal E\) in a frame rotating at \(\omega_L\).
    A weak seed pulse \(\varepsilon(t)\) initializes the first moments.
    The charger couples dissipatively to the battery mode \(b\), with resonance frequency \(\omega_b\) and damping rate \(\kappa_b\), through an auxiliary lossy mode \(c\), with resonance frequency \(\omega_c\) and damping rate \(\kappa_c\).
    The nonlocal reservoir channels with rate \(\Gamma\) generate the effective dissipative charger--battery interaction after adiabatic elimination of \(c\).
    \justifying}\label{Model}
\end{figure}

\textit{The Model}--
We now construct a fully Lindblad charger--battery model whose low-threshold instability arises from reservoir-engineered dissipation rather than from a direct coherent exchange interaction. Figure~\ref{Model} shows the schematic of the system. A harmonic oscillator with resonance frequency $\omega_a$ and local damping rate $\kappa_a$ plays the role of the charger. It interacts dissipatively with a second mode acting as the battery, characterized by a resonance frequency $\omega_b$ and a damping rate $\kappa_b$. The charger is parametrically driven by a classical two-photon field of physical frequency $2\omega_L$ and amplitude $\mathcal{E}$, which provides the energy ultimately stored in the battery. We assume that the charger and battery do not exchange energy through any direct coherent coupling. In a frame rotating at $\omega_L$, the dynamics is generated by a quadratic Hamiltonian together with linear Lindblad dissipators. The Hamiltonian is written as ($\hbar=1$)
\begin{equation}
\hat H(t) = \hat H_0 + \hat H_c + \hat H_{\rm int} + \hat H_{\rm pump} + \hat H_{\rm seed}(t),
\label{eq:H_general_split}
\end{equation}
where $\hat H_0 = \delta_a \hat a^\dagger \hat a + \delta_b \hat b^\dagger \hat b$ contains the bare modes energies (or detunings) with $\delta_j=\omega_j-\omega_L$, and $\hat H_{\rm pump} = \mathcal E\,\hat a^{\dagger 2} + \mathcal E^\ast \hat a^2$ is the single-mode two-photon pump acting solely on mode \(\hat a\). 
The short seed pulse, with form $\hat H_{\rm seed}(t) = \varepsilon(t)\,(\hat a +\, \hat a^\dagger)$ and amplitude $\varepsilon(t) = \varepsilon_0 \exp\!\left[-\frac{(t-t_0)^2}{2\sigma^2}\right]$, is kept weak and is used to initialize the first moments. 
Notice that the two-photon pump is quadratic: it changes the homogeneous drift matrix and can amplify fluctuations, but it does not displace an initial vacuum state. Hence, in the absence of the seed, \(|\langle \hat b(t)\rangle|^2\) remains null for all $t$ even when second moments grow; the seed only provides the coherent component that is amplified in the broken regime (see Note 1 of the SM~\cite{SuppMat}).

The interaction between the charger and the battery is mediated by a dissipative auxiliary mode $c$, characterized by a resonance frequency $\omega_c$ and damping rate $\kappa_c$, with Hamiltonian $\hat{H}_c=\delta_c\hat{c}^\dagger \hat{c}$. This strongly overdamped mode acts as a controllable, engineered reservoir that induces a purely dissipative coupling between the two subsystems. As explained in the Supplementary Materials, this arises when both the charger and the battery are coupled to $c$ through two nonlocal reservoirs with coupling rate $\Gamma$, producing an effective dissipative interaction between them \cite{PhysRevX.5.021025}. From this point onward, unless explicitly stated otherwise, we use the symmetric detuning convention $\delta_a=\delta$, $\delta_b=-\delta$, and $\delta_c=0$.
\begin{figure*}[t]
    \centering
    \includegraphics[width=2\columnwidth]{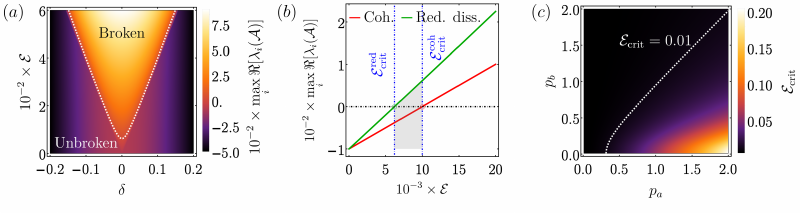}
    \caption{Stability thresholds of the reduced dissipative charger--battery model.
    (a) Stability indicator \(\max_i{\Re}[\lambda_i(\mathcal A(\mathcal E))]\) in the \((\delta,\mathcal E)\) plane, with \(\delta_a=-\delta_b=\delta\). The dotted contour marks the instability boundary \(\max_i{\Re}[\lambda_i]=0\).
    (b) Stability indicator versus pump amplitude for the reduced dissipative model and the coherent benchmark. For a fair comparison, the coherent coupling is chosen as \(|J|=|g|\), with \(g=\Gamma_{\rm eff}\mu_{ca}\mu_{cb}^\ast/2\). Vertical dotted lines and circle markers highlight \(\mathcal E_{\rm crit}^{\rm coh}\approx 1\times10^{-2}\) and \(\mathcal E_{\rm crit}^{\rm red}\approx 6.2\times10^{-3}\). The shaded gray area highlights the useful operating window with $\mathcal E_{\rm crit}^{\rm red}<|\mathcal E|<\mathcal E_{\rm crit}^{\rm coh}$.
    (c) Critical pump amplitude \(\mathcal E_{\rm crit}\) in the \((p_a,p_b)\) plane. The dotted contour marks \(\mathcal E_{\rm crit}=0.01\). Parameters are \(\delta_a=\delta_b=\delta_c=0\), \(\kappa_a=\kappa_b=0.02\), \(\kappa_c=0\), \(\Gamma=0.4\), \(p_a=1\), \(p_b=2\), \(p_{ca}=p_{cb}=10\), and \(J=0.2=|g|\).
    \justifying}\label{Fig2}
\end{figure*}

Assuming Markovian engineered reservoirs, the microscopic dissipative model is described by
\(
\dot{\hat\rho}
=
-i[\hat H_{3{\rm m}}(t),\hat\rho]
+
\kappa_a\mathcal D[\hat a]\hat\rho
+
\kappa_b\mathcal D[\hat b]\hat\rho
+
\kappa_c\mathcal D[\hat c]\hat\rho
+
\Gamma\mathcal D[\hat z_a]\hat\rho
+
\Gamma\mathcal D[\hat z_b]\hat\rho,
\)
where \(\mathcal D[\hat o]\hat\rho=\hat o\hat\rho\hat o^\dagger-\{\hat o^\dagger\hat o,\hat\rho\}/2\). Here \(\hat z_a=p_a\hat a+p_{ca}\hat c\) and \(\hat z_b=p_b\hat b+p_{cb}\hat c\) are the engineered nonlocal loss channels, and \(\hat H_{3{\rm m}}(t)=\delta\hat a^\dagger\hat a-\delta\hat b^\dagger\hat b+\mathcal E\hat a^{\dagger 2}+\mathcal E^\ast\hat a^2+\varepsilon(t)(\hat a+\hat a^\dagger)\). 
Thus, the charger and the battery are not coupled by a coherent beam-splitter Hamiltonian; their interaction is generated entirely by the shared dissipative channels. In the annihilation basis \(\hat d=(\hat a,\hat b,\hat c)^{\mathsf T}\), the dissipative part is equivalently specified by the positive Kossakowski matrix (see Note 2C of the SM \cite{SuppMat})
\begin{equation}
\mathsf K=
\begin{pmatrix}
\kappa_a+\Gamma_a & 0 & \Gamma\mu_{ca}\\
0 & \kappa_b+\Gamma_b & \Gamma\mu_{cb}\\
\Gamma\mu_{ca}^\ast & \Gamma\mu_{cb}^\ast & \mathsf K_{\rm ff}
\end{pmatrix},
\end{equation}
where \(\Gamma_a=\Gamma|p_a|^2\), \(\Gamma_b=\Gamma|p_b|^2\), \(\Gamma_{ca}=\Gamma|p_{ca}|^2\), \(\Gamma_{cb}=\Gamma|p_{cb}|^2\), \(\mu_{ca}=p_{ca}p_a^\ast\), \(\mu_{cb}=p_{cb}p_b^\ast\) and $\mathsf K_{\rm ff}\equiv\kappa_c+\Gamma_{ca}+\Gamma_{cb}$. 
Such dissipative couplings can be implemented using damped auxiliary cavities, waveguides, or transmission-line reservoirs \cite{Barzanjeh2017, PhysRevLett.123.127202, 2303.04358}. We focus on the Markovian limit; non-Markovian dissipative interactions provide a distinct route and are not needed for the mechanism studied here \cite{PRXQuantum.4.010306}.

In the fast-mediator regime, where $\mathsf K_{\rm ff}$ is the dominant relaxation scale, the auxiliary mode $\hat c$ can be eliminated at the Gaussian generator level. Partitioning $\mathsf K$ into the slow sector \((a,b)\) and the fast sector \(c\), the reduced two-mode dissipative kernel is the Schur complement
\begin{equation}
\mathsf K_{\rm red} =
\begin{pmatrix}
2\gamma_a & -2g\\
-2g^\ast & 2\gamma_b
\end{pmatrix},
\end{equation}
where
\(
\gamma_j = (\kappa_j+\Gamma_j-\Gamma_{\rm eff}|\mu_{cj}|^2)/2,
\)
\(
g = \Gamma_{\rm eff}\mu_{ca}\mu_{cb}^\ast/2,
\) and
\(
\Gamma_{\rm eff} = \Gamma^2/(\kappa_c+\Gamma_{ca}+\Gamma_{cb})
\).
Since the Schur complement of a positive Kossakowski matrix is again positive, the reduced model remains a completely positive Lindblad dynamics. Defining \(m^{(2)}=(\langle\hat a\rangle,\langle\hat b\rangle)^{\mathsf T}\) and \(\mathbf v^{(2)}=(m^{(2)},m^{(2)\ast})^{\mathsf T}\), the homogeneous first-moment dynamics takes the following form (see Note 2D of the SM \cite{SuppMat})
\begin{equation}
\dot{\mathbf v}^{(2)}(t)=\mathcal A^{(2)}_{\rm red}(\mathcal E)\mathbf v^{(2)}(t),
\quad
\mathcal A^{(2)}_{\rm red}(\mathcal E)=
\begin{pmatrix}
M^{(2)}_{\rm red} & Q(\mathcal E)\\
Q^\ast(\mathcal E) & M_{\rm red}^{(2)\ast}
\end{pmatrix},
\label{eq:bog_drift_general}
\end{equation}
with
\[
M^{(2)}_{\rm red} = 
\begin{pmatrix}
-\gamma_a-i\delta & g\\
g^\ast & -\gamma_b+i\delta
\end{pmatrix}.
\]
The rank-one pairing block \(Q(\mathcal E)\) is generated by the two-photon pump on the charger, with \(Q_{aa}=-2i\mathcal E\) as its only nonzero element. The unbroken regime is defined by all eigenvalues of \(\mathcal A^{(2)}_{\rm red}\) having negative real parts. The broken regime begins when
\(\max_i{\Re}[\lambda_i(\mathcal A^{(2)}_{\rm red})]>0\); then a weak seed has a component along an unstable mode, leading to exponential growth of the coherent battery response.

We stress that this stability criterion is a property of the controlled Markovian reduced drift matrix and must be interpreted on the parameter manifold inherited from the microscopic three-mode Lindblad model. It does not imply autonomous energy growth when the pump that supplies energy to the charger is absent. In that limit, and with no seed-induced displacement, the physically constrained three-mode drift matrix is stable, and all first moments relax to zero. The all-time validity conditions of the reduced drift matrix, together with direct comparisons to the exact three-mode dynamics, are given in Note~\ref{subsec:comparison_protocol} of the SM~\cite{SuppMat}.

This reduced dissipative model is the central object of the paper. The coherent two-mode interaction, obtained by replacing the dissipative off-diagonal entries \((g,\,g^\ast)\) in \(M_{\rm red}^{(2)}\) by the coherent beam-splitter entries \((-iJ,\,-iJ^\ast)\), with \(\hat H_{\rm int}^{\rm coh}=J\,\hat a^{\dagger}\hat b+J^{\ast}\hat a\hat b^{\dagger}\), is used only as a benchmark. The full three-mode dissipative model is used as a microscopic consistency check: in the fast-mediator regime, it reproduces the reduced dynamics, confirming that the enhancement is not an artifact of adiabatic elimination (see Note~2 of the SM~\cite{SuppMat}). The eliminated mediator leaves two signatures in the reduced model: a complex dissipative coupling \(g\), and renormalized damping rates \(\gamma_j\). The negative correction in \(\gamma_j\) ($\Gamma_{\rm eff}|\mu_{cj}|^2$) is the footprint of dissipative interference between the direct reservoir-loss channel and the pathway through the eliminated mediator mode $c$. Importantly, this correction does not make the passive two-mode dynamics unstable. Since the reduced Kossakowski matrix satisfies \(\mathsf K_{\rm red}\ge 0\), the passive drift matrix \(M_{\rm red}^{(2)}\) has no eigenvalue with positive real part when \(\mathcal E=0\) (see Note~2D of the SM~\cite{SuppMat}). Instead, the reservoir-engineered interference reduces the effective decay rate, or linewidth, of a collective charger--battery mode. We refer to this reduction of the relevant collective decay rate as linewidth softening. The term is used here in this specific dynamical sense: the passive decay eigenvalue is pushed closer to zero while remaining non-negative, similar collective-mode instabilities and soft-mode physics \cite{Mottl2012Science,Jakob2023NatCommun,Landig2015NatCommun,Cao2024Nature}. For \(\delta=0\), the smallest passive decay rate is \(\gamma_{\rm soft}=(\gamma_a+\gamma_b)/2-\sqrt{(\gamma_a-\gamma_b)^2/4+|g|^2}\), which remains non-negative because complete positivity imposes \(\gamma_a\gamma_b\ge |g|^2\). The advantage of the dissipative architecture is that this soft collective mode retains a finite battery component. The two-photon pump applied to the charger can therefore destabilize a low-linewidth collective mode while still transferring the amplified response (amplitude) to the battery. Thus, the threshold is not lowered by passive gain, but by linewidth softening through dissipative interference. In contrast, the coherent beam-splitter benchmark retains stronger reversible exchange dynamics and requires a larger pump amplitude to reach the broken regime.

Figure~\ref{Fig2} summarizes the stability mechanism behind the dissipative charging advantage. Fig.~\figpanel{Fig2}{a} shows the largest real part of the eigenvalues of the reduced dissipative model in the \((\delta,\mathcal E)\) plane. The dotted contour marks the parametric-instability boundary, \(\max_i{\Re}[\lambda_i(\mathcal A(\mathcal E))]=0\): outside this region the seeded response remains stable, while inside it one eigenmode acquires a positive growth rate. This boundary is not an exceptional point; rather, it is the pump-induced stability threshold of the drift matrix. The practical consequence is shown in Fig.~\figpanel{Fig2}{b}: the reservoir-engineered model reaches this threshold at a lower pump amplitude than the coherent beam-splitter benchmark. Importantly, this comparison is made at equal charger--battery coupling strength: the coherent benchmark is chosen with \(|J|=|g|\), where \(g=\Gamma_{\rm eff}\mu_{ca}\mu_{cb}^{\ast}/2\). Therefore, as mentioned above, the lower threshold is not a consequence of using a larger coupling in the dissipative model, but of reservoir-induced linewidth softening and dissipative interference. Since the resulting soft mode still has finite support on the battery's state, a pump applied to the charger can excite an exponentially growing battery response once the threshold is crossed. Fig.~\figpanel{Fig2}{c} demonstrates that this soft-mode-assisted threshold reduction is not confined to a single fine-tuned point, but persists over a broad region of reservoir couplings.
\begin{figure}[t]
    \centering
    \includegraphics[width=\columnwidth]{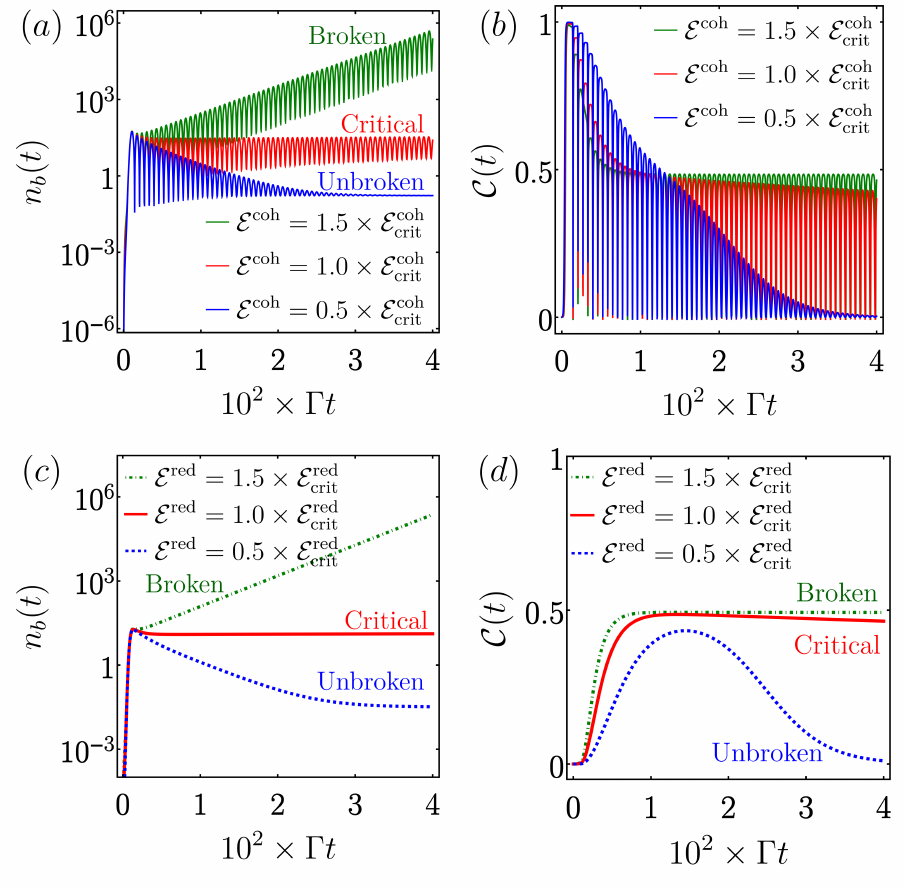}
    \caption{Time-domain charging below, at, and above threshold.
    (a,b) Coherent beam-splitter benchmark; (c,d) reduced dissipative model.
    Panels (a,c) show the battery occupation \(n_b(t)\) on a logarithmic scale, and panels (b,d) show the coherent fraction \(\mathcal{C}(t)\). For each architecture, the pump amplitudes are \(0.5\mathcal E_{\rm crit}\), \(\mathcal E_{\rm crit}\), and \(1.5\mathcal E_{\rm crit}\). Below threshold, the seeded response remains bounded, while above threshold \(n_b(t)\) grows exponentially. In the reduced dissipative model, the broken regime also yields the largest \(\mathcal{C}(t)\), showing that the growing occupation is dominated by seed-selected coherent charging rather than by incoherent fluctuation buildup. Parameters are the same as in Fig.~\ref{Fig2}, including the numerical values of $\mathcal E_{\rm crit}^{\rm coh}$ and $\mathcal E_{\rm crit}^{\rm red}$.
    \justifying}\label{Fig3}
\end{figure}

The spectral distinction in Fig.~\ref{Fig2} is directly reflected in the charging dynamics. Figure~\ref{Fig3} compares the battery occupation \(n_b(t)\) and the coherent fraction \(\mathcal{C}(t)\coloneqq|\langle \hat b(t)\rangle|^2/n_b(t)\) below, at, and above the corresponding threshold of each architecture. In both models, the seeded response remains bounded below threshold and grows exponentially above threshold. The coherent benchmark displays pronounced oscillations from reversible beam-splitter exchange, whereas the reduced dissipative model selects a cleaner growing mode.

The coherent fraction provides an operational diagnostic of this growth. Writing \(\delta\hat b(t) = \hat b(t) - \beta_b(t)\), with \(\beta_b(t)=\langle\hat b(t)\rangle\), one obtains
\(
n_b(t)=|\beta_b(t)|^2+\langle\delta\hat b^\dagger(t)\delta\hat b(t)\rangle .
\)
Thus \(\mathcal{C}(t)\) measures the fraction of the stored oscillator energy carried by the seed-selected displacement, rather than by fluctuation-induced occupation. Its enhancement in the broken dissipative regime shows that the instability does not merely heat the battery through parametric noise. Instead, the seed overlaps with the unstable soft mode, which retains a finite battery component, so the pump amplifies a phase-coherent battery displacement. More importantly, this coherent part can be directly recovered: applying a displacement \(D_b[-\beta_b(t)]\) extracts
\(
W_{\rm disp}(t)=\hbar\omega_b|\beta_b(t)|^2 ,
\)
giving a lower bound to the battery ergotropy. Therefore, the broken dissipative regime combines exponential charging with a large coherently addressable work fraction.

\begin{figure}[t]
    \centering
    \includegraphics[width=\columnwidth]{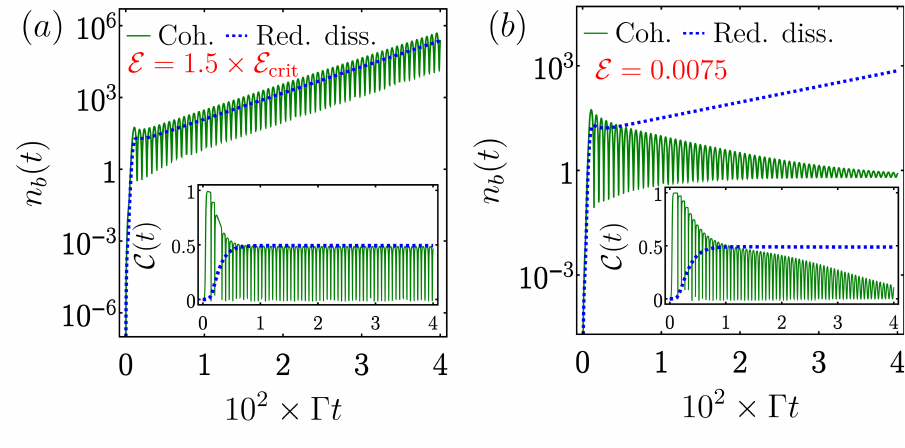}
    \caption{Direct dynamical comparison between the coherent benchmark and the reduced dissipative model (on a logarithmic scale).
    (a) Battery occupation at the same relative distance above threshold: \(\mathcal E^{\rm coh}=1.5\mathcal E_{\rm crit}^{\rm coh}\) and \(\mathcal E^{\rm red}=1.5\mathcal E_{\rm crit}^{\rm red}\).
    Both systems are in the broken regime, but the coherent model retains reversible exchange oscillations, whereas the dissipative model follows the reservoir-engineered growing mode.
    (b) Battery occupation at the same absolute pump amplitude, \(\mathcal E=0.0075\).
    At this drive, the reduced dissipative model is above threshold and charges exponentially, while the coherent benchmark remains below threshold.
    Insets show the coherent fraction \(\mathcal{C}(t)\).
    Parameters are the same as in Fig.~\ref{Fig2}.
    \justifying}\label{Fig4}
\end{figure}

Figure~\ref{Fig4} makes the threshold advantage explicit. At equal relative distance above threshold, both architectures enter the broken regime, but their dynamics remain qualitatively different. At the same weak absolute pump amplitude, however, only the reduced dissipative model is already above threshold, while the coherent benchmark remains effectively stable. Thus the reservoir-engineered interaction enables high-rate coherent charging at pump strengths for which the coherent architecture cannot yet exploit exponential growth.

\textit{Pump-power economy and coherent-work yield}--
The threshold reduction shown in Fig.~\ref{Fig2} has a direct experimental implication. The two-photon pump amplitude \(\mathcal E\) is a field amplitude, so the supplied pump power scales as \(|\mathcal E|^2\). As a result, the reduction of the instability threshold translates quadratically into a reduction of the critical pump power. For the parameters of Fig.~\ref{Fig2}, the reduced dissipative architecture reaches threshold at \(\mathcal E_{\rm crit}^{\rm red}\approx 6.2\times10^{-3}\), while the coherent benchmark reaches threshold at \(\mathcal E_{\rm crit}^{\rm coh}\approx 1.0\times10^{-2}\). This corresponds to a critical pump-power ratio \(\mathcal R_P=
\frac{P_{\rm crit}^{\rm red}}{P_{\rm crit}^{\rm coh}}
=
\left|
\frac{\mathcal E_{\rm crit}^{\rm red}}
{\mathcal E_{\rm crit}^{\rm coh}}
\right|^2\approx 0.39\), or about \(61\%\) less threshold pump power, equivalently a saving of approximately \(4.1\,{\rm dB}\) (see Note~3 of the SM~\cite{SuppMat}). Since this comparison is performed at equal effective charger--battery coupling strength, \(|J|=|g|\), the saving is not obtained by using a stronger interaction. It is instead a consequence of dissipative linewidth softening of the collective mode that is destabilized by the pump.

This power reduction opens a useful operating window. As shown by the shaded gray area in Panel (b) of Fig.~\ref{Fig2}, for pump amplitudes in the range \(\mathcal E_{\rm crit}^{\rm red}<|\mathcal E|<\mathcal E_{\rm crit}^{\rm coh}\), the reduced dissipative battery is already in the broken regime, whereas the coherent benchmark remains below threshold. Figure~\ref{Fig4} also demonstrates this situation at \(\mathcal E=0.0075\): the dissipative model charges exponentially, while the coherent benchmark gives only a bounded response. The reservoir-engineered interaction, therefore, does more than shift a spectral boundary. It makes fast charging accessible at pump powers for which a fair coherent architecture cannot yet exploit exponential growth.

The advantage can also be stated as a reduction of pump action, namely the time-integrated pump power required to reach a target battery energy. For a target occupation \(N_\star\), this cost is proportional to \(\int_0^{t_\star}|\mathcal E(t)|^2dt\), with \(n_b(t_\star)=N_\star\). More operationally, the target can be chosen as the coherently addressable work \(W_{\rm disp}(t)=\hbar\omega_b|\langle\hat b(t)\rangle|^2\).
The reduced dissipative model is favorable for this metric since the same broken regime that lowers the threshold also yields a large coherent fraction, as shown in Fig.~\ref{Fig3}. Thus, the amplified battery occupation is not merely fluctuation-induced heating, but contains a large seed-selected displacement component that is directly addressable as useful work.

This gives the mechanism a hardware-facing interpretation: reservoir engineering does not supply the stored energy, but it reduces the pump power required for the external drive to store that energy in a coherent, phase-addressable form. This distinction is important for superconducting, optomechanical, and magnonic platforms, where parametric pumps are available, but pump power, saturation, and cryogenic heat load are practical constraints \cite{Krinner2019,Hougland2025,Kurman2026,ahmadi2026LandauZener,ahmadi2026Amplifier}. The present architecture therefore provides a route to pump-efficient coherent quantum energy for storage: the dissipative mediator lowers the critical pump power while the unstable soft mode retains enough battery's state component to convert the reduced pump cost into coherently accessible stored work.

The occupation plotted in Figs.~\ref{Fig3} and~\ref{Fig4} corresponds to the stored oscillator energy scale, $E_b(t)=\hbar\omega_b n_b(t)$, up to the usual rotating-frame convention. We also show the coherent fraction $\mathcal{C}(t)$ to separate seed-selected coherent charging from fluctuation-induced occupation; a full ergotropy and discharge optimization can then be built on top of the same Gaussian solution. This operational distinction is important because the threshold reduction is not a hidden passive gain, but dissipative linewidth engineering. In this sense, the device acts as a pump-efficient, seed-triggered quantum energy buffer: below threshold, the battery response is bounded, whereas above threshold, the weak seed selects an amplified storage mode. The application window is clearest in superconducting and optomechanical architectures, where reservoir-engineered parametric devices already use dissipative interactions for quantum-limited amplification \cite{PhysRevLett.112.133904,PhysRevX.5.021025} and where pump power, saturation, and cryogenic heat load are practical constraints \cite{Hougland2025}. The same physical principle also parallels anti-PT wireless-power circuits, where dissipative/coherent-coupling competition produces level pinning and robust energy transfer \cite{Guo2024LevelPinningWPT}. Finally, recent microcavity quantum-battery experiments, including a complete charge--discharge cycle with superextensive electrical output, indicate that rapid quantum charging is becoming a practical energy-conversion problem rather than a purely theoretical issue~\cite{eabk3160,Hymas2026SuperextensiveQB}.

\noindent

\textit{Summary--}.
We have proposed a reservoir-engineered quantum battery in which a passive dissipative mediator lowers the pump threshold for entering a broken dynamical regime. The key mechanism is linewidth softening through dissipative interference: the eliminated mediator creates a low-decay collective mode that remains coupled to the battery, while complete positivity keeps the passive dynamics stable in the absence of the two-photon pump. The external pump then destabilizes this soft mode at a lower critical amplitude than in a coherent beam-splitter charger--battery benchmark, producing high-rate seeded charging with no oscillations. Beyond fast growth, the broken dissipative regime also yields a large coherent fraction, showing that the amplified occupation is mainly stored as a seed-selected battery displacement rather than as incoherent fluctuation-induced energy. This coherently addressable component can be extracted by a displacement operation on the battery and therefore provides a direct lower bound on the available ergotropy. The construction remains rooted in a microscopic three-mode Lindblad model and is validated against the full dynamics in the fast-mediator regime.

\textit{Acknowledgements}--
BA acknowledges support by Polish National Agency for Academic Exchange under the Strategic Partnership Programme grant BNI/PST/2023/1/00013/U/00001. PH acknowledge support from IRA Programme (project no. FENG.02.01-IP.05-0006/23) financed by the FENG program 2021-2027, Priority FENG.02, Measure FENG.02.01., with the support of the FNP. A.H.A.M. acknowledges support from National Science Centre, Poland Grant OPUS-21 (No. 2021/41/B/ST2/03207). PM acknowledges support by the Polish National Agency for Academic Exchange (NAWA), under Strategic Partnerships Programme, project number BNI/PST/2023/1/00013/U/00001. S.B. acknowledges funding by the Natural Sciences and Engineering Research Council of Canada (NSERC) through its Discovery Grant.

\bibliography{References}

@misc{SuppMat,
  author    = {},
  title     = {},
  year      = {},
  note      = {See Supplemental Material for the derivation of equations of motion and the adiabatic elimination.}
}

@article{Beaulieu2025NatCommun,
  author  = {Beaulieu, Guillaume and Minganti, Fabrizio and Frasca, Simone and Savona, Vincenzo and Felicetti, Simone and Di Candia, Roberto and Scarlino, Pasquale},
  title   = {Observation of first- and second-order dissipative phase transitions in a two-photon driven Kerr resonator},
  journal = {Nature Communications},
  volume  = {16},
  pages   = {1954},
  year    = {2025},
  doi     = {10.1038/s41467-025-56830-w}
}

@article{PhysRevE.102.052109,
  title = {Entanglement, coherence, and charging process of quantum batteries},
  author = {Kamin, F. H. and Tabesh, F. T. and Salimi, S. and Santos, Alan C.},
  journal = {Phys. Rev. E},
  volume = {102},
  issue = {5},
  pages = {052109},
  numpages = {7},
  year = {2020},
  month = {Nov},
  publisher = {American Physical Society},
  doi = {10.1103/PhysRevE.102.052109},
  url = {https://link.aps.org/doi/10.1103/PhysRevE.102.052109}
}

@article{Mottl2012Science,
  author  = {Mottl, Rafael and Brennecke, Ferdinand and Baumann, Kristian and Landig, Renate and Donner, Tobias and Esslinger, Tilman},
  title   = {Roton-Type Mode Softening in a Quantum Gas with Cavity-Mediated Long-Range Interactions},
  journal = {Science},
  volume  = {336},
  number  = {6088},
  pages   = {1570--1573},
  year    = {2012},
  doi     = {10.1126/science.1220314}
}

@article{Landig2015NatCommun,
  author  = {Landig, Renate and Brennecke, Ferdinand and Mottl, Rafael and Donner, Tobias and Esslinger, Tilman},
  title   = {Measuring the Dynamic Structure Factor of a Quantum Gas Undergoing a Structural Phase Transition},
  journal = {Nature Communications},
  volume  = {6},
  pages   = {7046},
  year    = {2015},
  doi     = {10.1038/ncomms8046}
}

@article{Jakob2023NatCommun,
  author  = {Jakob, Lukas A. and Deacon, William M. and Zhang, Yuan and de Nijs, Bart and Pavlenko, Elena and Hu, Shu and Carnegie, Cloudy and Neuman, Tomas and Esteban, Ruben and Aizpurua, Javier and Baumberg, Jeremy J.},
  title   = {Giant Optomechanical Spring Effect in Plasmonic Nano- and Picocavities Probed by Surface-Enhanced Raman Scattering},
  journal = {Nature Communications},
  volume  = {14},
  pages   = {3291},
  year    = {2023},
  doi     = {10.1038/s41467-023-38124-1}
}

@article{Cao2024Nature,
  author  = {Cao, Ruyue and Yang, Qiao-Lin and Deng, Hui-Xiong and Wei, Su-Huai and Robertson, John and Luo, Jun-Wei},
  title   = {Softening of the Optical Phonon by Reduced Interatomic Bonding Strength without Depolarization},
  journal = {Nature},
  volume  = {634},
  pages   = {1080--1085},
  year    = {2024},
  doi     = {10.1038/s41586-024-08099-0}
}

@article{Krantz2016NatCommun,
  author  = {Krantz, Philip and Bengtsson, Andreas and Simoen, Micha{\"e}l and Gustavsson, Simon and Oliver, W. D. and Wilson, C. M. and Delsing, Per and Bylander, Jonas},
  title   = {Single-shot read-out of a superconducting qubit using a Josephson parametric oscillator},
  journal = {Nature Communications},
  volume  = {7},
  pages   = {11417},
  year    = {2016},
  doi     = {10.1038/ncomms11417}
}

@book{Lyapunov1992Stability,
  author    = {A. M. Lyapunov},
  title     = {The General Problem of the Stability of Motion},
  publisher = {Taylor and Francis},
  year      = {1992},
  url={https://api.semanticscholar.org/CorpusID:119995127}
}

@book{Khalil2002NonlinearSystems,
  author    = {Hassan K. Khalil},
  title     = {Nonlinear Systems},
  edition   = {3},
  publisher = {Prentice Hall},
  year      = {2002},
  url={https://api.semanticscholar.org/CorpusID:59790953}
}

@book{Trefethen2005Spectra,
  author    = {Lloyd N. Trefethen and Mark Embree},
  title     = {Spectra and Pseudospectra: The Behavior of Nonnormal Matrices and Operators},
  publisher = {Princeton University Press},
  year      = {2005},
  url       = {https://doi.org/10.2307/j.ctvzxx9kj}
}

@article{Poyatos1996ReservoirEngineering,
  author  = {J. F. Poyatos and J. I. Cirac and P. Zoller},
  title   = {Quantum Reservoir Engineering with Laser Cooled Trapped Ions},
  journal = {Phys. Rev. Lett.},
  volume  = {77},
  pages   = {4728},
  year    = {1996},
  doi = {https://doi.org/10.1103/PhysRevLett.77.4728}
}

@article{Verstraete2009Dissipation,
  author  = {F. Verstraete and M. M. Wolf and J. I. Cirac},
  title   = {Quantum computation and quantum-state engineering driven by dissipation},
  journal = {Nature Physics},
  volume  = {5},
  pages   = {633--636},
  year    = {2009},
  doi = {https://doi.org/10.1038/nphys1342}
}

@article{ferraro2026opportunities,
  title={Opportunities and challenges of quantum batteries},
  author={Ferraro, Dario and Cavaliere, Fabio and Genoni, Marco G and Benenti, Giuliano and Sassetti, Maura},
  journal={Nature Reviews Physics},
  pages={115–127},
  volume={8},
  year={2026},
  publisher={Nature Publishing Group},
  url={https://doi.org/10.1038/s42254-025-00906-5}
}

@article{Ashida2020NonHermitianReview,
  title        = {Non-Hermitian physics},
  author       = {Ashida, Yuto and Gong, Zongping and Ueda, Masahito},
  journal      = {Advances in Physics},
  volume       = {69},
  number       = {3},
  pages        = {249--435},
  year         = {2020},
  doi          = {10.1080/00018732.2021.1876991}
}

@article{ElGanainy2018NonHermitian,
  title        = {Non-Hermitian physics and {PT} symmetry},
  author       = {El-Ganainy, Ramy and Makris, Konstantinos G. and Khajavikhan, Mercedeh and Musslimani, Ziad H. and Rotter, Stefan and Christodoulides, Demetrios N.},
  journal      = {Nature Physics},
  volume       = {14},
  pages        = {11--19},
  year         = {2018},
  doi          = {10.1038/nphys4323}
}

@article{Ozdemir2019EPReview,
  title        = {Parity--time symmetry and exceptional points in photonics},
  author       = {{\"O}zdemir, {\c{S}}ahin Kaya and Rotter, Stefan and Nori, Franco and Yang, Lan},
  journal      = {Nature Materials},
  volume       = {18},
  pages        = {783--798},
  year         = {2019},
  doi          = {10.1038/s41563-019-0304-9}
}

@article{Ruter2010PTSymmetry,
  title        = {Observation of parity--time symmetry in optics},
  author       = {R{\"u}ter, C. E. and Makris, K. G. and El-Ganainy, R. and Christodoulides, D. N. and Segev, M. and Kip, D.},
  journal      = {Nature Physics},
  volume       = {6},
  pages        = {192--195},
  year         = {2010},
  doi          = {10.1038/nphys1515}
}

@article{Feng2013Unidirectional,
  title        = {Experimental demonstration of a unidirectional reflectionless parity--time metamaterial at optical frequencies},
  author       = {Feng, Liang and Xu, Yi-Lin and Fegadolli, William S. and Lu, Ming-Hui and Oliveira, Jo{\~a}o E. B. and Almeida, Vilson R. and Chen, Yan-Feng and Scherer, Axel},
  journal      = {Nature Materials},
  volume       = {12},
  pages        = {108--113},
  year         = {2013},
  doi          = {10.1038/nmat3495}
}

@article{Feng2014SingleModeLaser,
  title        = {Single-mode laser by parity-time symmetry breaking},
  author       = {Feng, Liang and Wong, Zi Jing and Ma, Ren-Min and Wang, Yuan Wang and Zhang, Xiang},
  journal      = {Science},
  volume       = {346},
  number       = {6212},
  pages        = {972--975},
  year         = {2014},
  doi          = {10.1126/science.1258479}
}

@article{Hodaei2014SingleModeLaser,
  title        = {Parity-time--symmetric microring lasers},
  author       = {Hodaei, Hossein and Miri, Mohammad-Ali and Heinrich, Matthias and Christodoulides, Demetrios N. and Khajavikhan, Mercedeh},
  journal      = {Science},
  volume       = {346},
  number       = {6212},
  pages        = {975--978},
  year         = {2014},
  doi          = {10.1126/science.1258480}
}

@article{Naghiloo2019QuantumEP,
  title        = {Quantum state tomography across the exceptional point in a single dissipative qubit},
  author       = {Naghiloo, M. and Abbasi, M. and Joglekar, Y. N. and Murch, K. W.},
  journal      = {Nature Physics},
  volume       = {15},
  pages        = {1232--1236},
  year         = {2019},
  doi          = {10.1038/s41567-019-0652-z}
}

@book{Roger1985,
  title={Matrix Analysis},
  author={Roger A. Horn and Charles R. Johnson},
  year={1985},
  publisher={Cambridge University Press},
  url={https://doi.org/10.1017/CBO9780511810817}
}

@article{PhysRevLett.120.117702,
  title = {High-Power Collective Charging of a Solid-State Quantum Battery},
  author = {Ferraro, Dario and Campisi, Michele and Andolina, Gian Marcello and Pellegrini, Vittorio and Polini, Marco},
  journal = {Phys. Rev. Lett.},
  volume = {120},
  issue = {11},
  pages = {117702},
  numpages = {6},
  year = {2018},
  month = {Mar},
  publisher = {American Physical Society},
  doi = {10.1103/PhysRevLett.120.117702},
  url = {https://link.aps.org/doi/10.1103/PhysRevLett.120.117702}
}

@article{PhysRevA.107.042419,
  title = {Catalysis in charging quantum batteries},
  author = {Rodr\'{\i}guez, R. R. and Ahmadi, B. and Mazurek, P. and Barzanjeh, S. and Alicki, R. and Horodecki, P.},
  journal = {Phys. Rev. A},
  volume = {107},
  issue = {4},
  pages = {042419},
  numpages = {8},
  year = {2023},
  month = {Apr},
  publisher = {American Physical Society},
  doi = {10.1103/PhysRevA.107.042419},
  url = {https://link.aps.org/doi/10.1103/PhysRevA.107.042419}
}

@article{8xsm-5mb6,
  title = {Wireless energy transfer in a non-Hermitian quantum battery},
  author = {Yang, Fang-Mei and Dou, Fu-Quan},
  journal = {Phys. Rev. A},
  volume = {112},
  issue = {4},
  pages = {042205},
  numpages = {11},
  year = {2025},
  month = {Oct},
  publisher = {American Physical Society},
  doi = {10.1103/8xsm-5mb6},
  url = {https://link.aps.org/doi/10.1103/8xsm-5mb6}
}

@article{PhysRevE.87.042123,
  title = {Entanglement boost for extractable work from ensembles of quantum batteries},
  author = {Alicki, Robert and Fannes, Mark},
  journal = {Phys. Rev. E},
  volume = {87},
  issue = {4},
  pages = {042123},
  numpages = {4},
  year = {2013},
  month = {Apr},
  publisher = {American Physical Society},
  doi = {10.1103/PhysRevE.87.042123},
  url = {https://link.aps.org/doi/10.1103/PhysRevE.87.042123}
}

@article{PhysRevResearch.5.013155,
  title = {Quantum advantage in charging cavity and spin batteries by repeated interactions},
  author = {Salvia, Raffaele and Perarnau-Llobet, Mart\'{\i} and Haack, G\'eraldine and Brunner, Nicolas and Nimmrichter, Stefan},
  journal = {Phys. Rev. Res.},
  volume = {5},
  issue = {1},
  pages = {013155},
  numpages = {13},
  year = {2023},
  month = {Feb},
  publisher = {American Physical Society},
  doi = {10.1103/PhysRevResearch.5.013155},
  url = {https://link.aps.org/doi/10.1103/PhysRevResearch.5.013155}
}

@article{ll76-j2l5,
  title = {Essay: Topological Phases and Exceptional Points in Non-Hermitian Systems},
  author = {Xue, Peng},
  journal = {Phys. Rev. Lett.},
  volume = {136},
  issue = {17},
  pages = {170001},
  numpages = {11},
  year = {2026},
  month = {Apr},
  publisher = {American Physical Society},
  doi = {10.1103/ll76-j2l5},
  url = {https://link.aps.org/doi/10.1103/ll76-j2l5}
}

@article{ahmadi2026ChiralSqueezing,
  title={Charging Quantum Batteries with Chiral Squeezing},
  author={Ahmadi, Borhan and H. A. Malavazi, André and Splettstoesser, Janine and Horodecki, Paweł and Du, Lei},
  journal={	arXiv:2606.16764},
  year={2026},
  url={https://doi.org/10.48550/arXiv.2606.16764}
}

@article{Rodriguez_2024,
    doi = {10.1088/1367-2630/ad3843},
    url = {https://dx.doi.org/10.1088/1367-2630/ad3843},
    year = {2024},
    month = {apr},
    publisher = {IOP Publishing},
    volume = {26},
    number = {4},
    pages = {043004},
    author = {R R Rodríguez and B Ahmadi and G Suárez and P Mazurek and S Barzanjeh and P Horodecki},
    title = {{Optimal quantum control of charging quantum batteries}},
    journal = {New Journal of Physics}
}

@article{PhysRevLett.123.127202,
  title = {Nonreciprocity and Unidirectional Invisibility in Cavity Magnonics},
  author = {Wang, Yi-Pu and Rao, J. W. and Yang, Y. and Xu, Peng-Chao and Gui, Y. S. and Yao, B. M. and You, J. Q. and Hu, C.-M.},
  journal = {Phys. Rev. Lett.},
  volume = {123},
  issue = {12},
  pages = {127202},
  numpages = {6},
  year = {2019},
  month = {Sep},
  publisher = {American Physical Society},
  doi = {10.1103/PhysRevLett.123.127202},
  url = {https://link.aps.org/doi/10.1103/PhysRevLett.123.127202}
}

@article{2303.04358,
  title = {Nonreciprocity in cavity magnonics at millikelvin temperature},
  author = {Kim, Mun and Tabesh, Armin and Zegray, Tyler and Barzanjeh, Shabir and Hu, Can-Ming},
  journal = {Journal of Applied Physics},
  volume = {135},
  number = {6},
  pages = {063904},
  year = {2024},
  publisher = {AIP Publishing},
  doi = {10.1063/5.0176462},
  url = {https://doi.org/10.1063/5.0176462}
}

@article{eabk3160,
author = {James Q. Quach  and Kirsty E. McGhee  and Lucia Ganzer  and Dominic M. Rouse  and Brendon W. Lovett  and Erik M. Gauger  and Jonathan Keeling  and Giulio Cerullo  and David G. Lidzey  and Tersilla Virgili },
title = {Superabsorption in an organic microcavity: Toward a quantum battery},
journal = {Science Advances},
volume = {8},
number = {2},
pages = {3160},
year = {2022},
doi = {10.1126/sciadv.abk3160},
URL = {https://www.science.org/doi/abs/10.1126/sciadv.abk3160}}

@article{PhysRevApplied.14.024092,
  title = {Using Dark States to Charge and Stabilize Open Quantum Batteries},
  author = {Quach, James Q. and Munro, William J.},
  journal = {Phys. Rev. Appl.},
  volume = {14},
  issue = {2},
  pages = {024092},
  numpages = {9},
  year = {2020},
  month = {Aug},
  publisher = {American Physical Society},
  doi = {10.1103/PhysRevApplied.14.024092},
  url = {https://link.aps.org/doi/10.1103/PhysRevApplied.14.024092}
}

@Article{Barzanjeh2017,
author={Barzanjeh, S.
and Wulf, M.
and Peruzzo, M.
and Kalaee, M.
and Dieterle, P. B.
and Painter, O.
and Fink, J. M.},
title={Mechanical on-chip microwave circulator},
journal={Nature Communications},
year={2017},
month={Oct},
day={16},
volume={8},
number={1},
pages={953},
issn={2041-1723},
doi={10.1038/s41467-017-01304-x},
url={https://doi.org/10.1038/s41467-017-01304-x}
}

@article{PhysRevApplied.4.034002,
  title = {On-Chip Superconducting Microwave Circulator from Synthetic Rotation},
  author = {Kerckhoff, Joseph and Lalumi\`ere, Kevin and Chapman, Benjamin J. and Blais, Alexandre and Lehnert, K. W.},
  journal = {Phys. Rev. Appl.},
  volume = {4},
  issue = {3},
  pages = {034002},
  numpages = {14},
  year = {2015},
  month = {Sep},
  publisher = {American Physical Society},
  doi = {10.1103/PhysRevApplied.4.034002},
  url = {https://link.aps.org/doi/10.1103/PhysRevApplied.4.034002}
}

@article{PhysRevX.7.041043,
  title = {Widely Tunable On-Chip Microwave Circulator for Superconducting Quantum Circuits},
  author = {Chapman, Benjamin J. and Rosenthal, Eric I. and Kerckhoff, Joseph and Moores, Bradley A. and Vale, Leila R. and Mates, J. A. B. and Hilton, Gene C. and Lalumi\`ere, Kevin and Blais, Alexandre and Lehnert, K. W.},
  journal = {Phys. Rev. X},
  volume = {7},
  issue = {4},
  pages = {041043},
  numpages = {16},
  year = {2017},
  month = {Nov},
  publisher = {American Physical Society},
  doi = {10.1103/PhysRevX.7.041043},
  url = {https://link.aps.org/doi/10.1103/PhysRevX.7.041043}
}

@article{PhysRevLett.112.133904,
  title = {Quantum-Limited Amplification via Reservoir Engineering},
  author = {Metelmann, A. and Clerk, A. A.},
  journal = {Phys. Rev. Lett.},
  volume = {112},
  issue = {13},
  pages = {133904},
  numpages = {5},
  year = {2014},
  month = {Apr},
  publisher = {American Physical Society},
  doi = {10.1103/PhysRevLett.112.133904},
  url = {https://link.aps.org/doi/10.1103/PhysRevLett.112.133904}
}

@article{ahmadi2026LandauZener,
  title={Quantum-Battery-Powered Geometric Landau-Zener Interferometry},
  author={Ahmadi, Borhan},
  journal={arXiv:2605.18108},
  year={2026},
  url={https://doi.org/10.48550/arXiv.2605.18108}
}

@article{ahmadi2026Amplifier,
  title={Quantum Batteries as Work Sources for Phase-Locked Parametric Amplification},
  author={Ahmadi, Borhan},
  journal={arXiv:2606.20306},
  year={2026},
  url={https://doi.org/10.48550/arXiv.2606.20306}
}

@ARTICLE{Toth2017,
  author = {Toth, L. D. and Bernier, N. R. and Nunnenkamp, A. and Feofanov, A.
	K. and Kippenberg, T. J.},
  title = {A dissipative quantum reservoir for microwave light using a mechanical
	oscillator},
  journal = {Nat Phys},
  year = {2017},
  volume = {13},
  pages = {787-793},
  month = may,
  issn = {1745-2481},
  publisher = {Nature Publishing Group},
  url = {http://dx.doi.org/10.1038/nphys4121}
}

@article{PhysRevA.108.022215,
  title = {Quantum parameter estimation of non-Hermitian systems with optimal measurements},
  author = {Yu, Xinglei and Zhang, Chengjie},
  journal = {Phys. Rev. A},
  volume = {108},
  issue = {2},
  pages = {022215},
  numpages = {12},
  year = {2023},
  month = {Aug},
  publisher = {American Physical Society},
  doi = {10.1103/PhysRevA.108.022215},
  url = {https://link.aps.org/doi/10.1103/PhysRevA.108.022215}
}

@article{Krinner2019,
  author  = {Krinner, S. and Storz, S. and Kurpiers, P. and Magnard, P. and Heinsoo, J. and Keller, R. and Luetolf, J. and Eichler, C. and Wallraff, A.},
  title   = {Engineering cryogenic setups for 100-qubit scale superconducting circuit systems},
  journal = {EPJ Quantum Technology},
  volume  = {6},
  pages   = {2},
  year    = {2019},
  url     = {https://doi.org/10.1140/epjqt/s40507-019-0072-0}
}

@article{Hougland2025,
  title = {Pump-efficient Josephson parametric amplifiers with high saturation power},
  author = {Hougland, Nicholas M. and Li, Zhuan and Kaufman, Ryan and Mesits, Boris and Mong, Roger S. K. and Hatridge, Michael and Pekker, David},
  journal = {Phys. Rev. A},
  volume = {111},
  issue = {2},
  pages = {022611},
  year = {2025},
  month = {Feb},
  publisher = {American Physical Society},
  doi = {10.1103/PhysRevA.111.022611},
  url = {https://link.aps.org/doi/10.1103/PhysRevA.111.022611}
}

@article{Kurman2026,
  title = {Powering Quantum Computation with Quantum Batteries},
  author = {Kurman, Yaniv and Hymas, Kieran and Fedorov, Arkady and Munro, William J. and Quach, James},
  journal = {Phys. Rev. X},
  volume = {16},
  issue = {1},
  pages = {011016},
  numpages = {17},
  year = {2026},
  month = {Jan},
  publisher = {American Physical Society},
  doi = {10.1103/l39v-jwwz},
  url = {https://link.aps.org/doi/10.1103/l39v-jwwz}
}

@article{PhysRevLett.131.030402,
  title = {Battery Capacity of Energy-Storing Quantum Systems},
  author = {Yang, Xue and Yang, Yan-Han and Alimuddin, Mir and Salvia, Raffaele and Fei, Shao-Ming and Zhao, Li-Ming and Nimmrichter, Stefan and Luo, Ming-Xing},
  journal = {Phys. Rev. Lett.},
  volume = {131},
  issue = {3},
  pages = {030402},
  numpages = {7},
  year = {2023},
  month = {Jul},
  publisher = {American Physical Society},
  doi = {10.1103/PhysRevLett.131.030402},
  url = {https://link.aps.org/doi/10.1103/PhysRevLett.131.030402}
}

@article{PRXQuantum.4.010306,
  title = {Quantum Nonreciprocal Interactions via Dissipative Gauge Symmetry},
  author = {Wang, Yu-Xin and Wang, Chen and Clerk, Aashish A.},
  journal = {PRX Quantum},
  volume = {4},
  issue = {1},
  pages = {010306},
  numpages = {28},
  year = {2023},
  month = {Jan},
  publisher = {American Physical Society},
  doi = {10.1103/PRXQuantum.4.010306},
  url = {https://link.aps.org/doi/10.1103/PRXQuantum.4.010306}
}

@article{PhysRevApplied.23.024010,
  title = {Superoptimal charging of quantum batteries via reservoir engineering: Arbitrary energy transfer unlocked},
  author = {Ahmadi, Borhan and Mazurek, Pawe\l{} and Barzanjeh, Shabir and Horodecki, Pawe\l{}},
  journal = {Phys. Rev. Appl.},
  volume = {23},
  issue = {2},
  pages = {024010},
  numpages = {14},
  year = {2025},
  month = {Feb},
  publisher = {American Physical Society},
  doi = {10.1103/PhysRevApplied.23.024010},
  url = {https://link.aps.org/doi/10.1103/PhysRevApplied.23.024010}
}

@article{ahmadi2025harnessing,
  title={Harnessing Environmental Noise for Quantum Energy Storage},
  author={Ahmadi, Borhan and Ravichandran, Aravinth Balaji and Mazurek, Pawe{\l} and Barzanjeh, Shabir and Horodecki, Pawe{\l}},
  journal={arXiv:2510.06384},
  year={2025},
  url={https://doi.org/10.48550/arXiv.2510.06384}
}

@article{PhysRevX.5.021025,
  title = {Nonreciprocal Photon Transmission and Amplification via Reservoir Engineering},
  author = {Metelmann, A. and Clerk, A. A.},
  journal = {Phys. Rev. X},
  volume = {5},
  issue = {2},
  pages = {021025},
  numpages = {16},
  year = {2015},
  month = {Jun},
  publisher = {American Physical Society},
  doi = {10.1103/PhysRevX.5.021025},
  url = {https://link.aps.org/doi/10.1103/PhysRevX.5.021025}
}

@article{6kwv-z6fx,
  title = {Reliable quantum advantage in quantum battery charging},
  author = {Rinaldi, Davide and Filip, Radim and Gerace, Dario and Guarnieri, Giacomo},
  journal = {Phys. Rev. A},
  volume = {112},
  issue = {1},
  pages = {012205},
  numpages = {11},
  year = {2025},
  month = {Jul},
  publisher = {American Physical Society},
  doi = {10.1103/6kwv-z6fx},
  url = {https://link.aps.org/doi/10.1103/6kwv-z6fx}
}

@article{PhysRevLett2019,
  title = {Perfectly Absorbing Exceptional Points and Chiral Absorbers},
  author = {Sweeney, William R. and Hsu, Chia Wei and Rotter, Stefan and Stone, A. Douglas},
  journal = {Phys. Rev. Lett.},
  volume = {122},
  issue = {9},
  pages = {093901},
  numpages = {6},
  year = {2019},
  month = {Mar},
  publisher = {American Physical Society},
  doi = {10.1103/PhysRevLett.122.093901}
}

@article{
changqing2021science,
author = {Changqing Wang  and William R. Sweeney  and A. Douglas Stone  and Lan Yang },
title = {Coherent perfect absorption at an exceptional point},
journal = {Science},
volume = {373},
number = {6560},
pages = {1261-1265},
year = {2021},
doi = {10.1126/science.abj1028}}

@article{PhysRevLett.134.180401,
  title = {Topological Quantum Batteries},
  author = {Lu, Zhi-Guang and Tian, Guoqing and L\"u, Xin-You and Shang, Cheng},
  journal = {Phys. Rev. Lett.},
  volume = {134},
  issue = {18},
  pages = {180401},
  numpages = {8},
  year = {2025},
  month = {May},
  publisher = {American Physical Society},
  doi = {10.1103/PhysRevLett.134.180401},
  url = {https://link.aps.org/doi/10.1103/PhysRevLett.134.180401}
}

@article{camposeo2025quantum,
  title={Quantum Batteries: A Materials Science Perspective},
  author={Camposeo, Andrea and Virgili, Tersilla and Lombardi, Floriana and Cerullo, Giulio and Pisignano, Dario and Polini, Marco},
  journal={Advanced Materials},
  issue = {17},
  volume = {37},
  pages={2415073},
  year={2025},
  publisher={Wiley Online Library},
  url={https://doi.org/10.1002/adma.202415073}
}

@article{zhang2024quantum,
  title={Quantum Battery in the Heisenberg Spin Chain Models with Dzyaloshinskii-Moriya Interaction},
  author={Zhang, Xiang-Long and Song, Xue-Ke and Wang, Dong},
  journal={Advanced Quantum Technologies},
  volume={7},
  number={9},
  pages={2400114},
  year={2024},
  publisher={Wiley Online Library},
  url = {https://doi.org/10.1002/qute.202400114}
}

@article{PhysRevA.109.042207,
  title = {Quantum battery with non-Hermitian charging},
  author = {Konar, Tanoy Kanti and Lakkaraju, Leela Ganesh Chandra and Sen (De), Aditi},
  journal = {Phys. Rev. A},
  volume = {109},
  issue = {4},
  pages = {042207},
  numpages = {12},
  year = {2024},
  month = {Apr},
  publisher = {American Physical Society},
  doi = {10.1103/PhysRevA.109.042207},
  url = {https://link.aps.org/doi/10.1103/PhysRevA.109.042207}
}

@article{PhysRevLett.132.210402,
  title = {Nonreciprocal Quantum Batteries},
  author = {Ahmadi, B. and Mazurek, P. and Horodecki, P. and Barzanjeh, S.},
  journal = {Phys. Rev. Lett.},
  volume = {132},
  issue = {21},
  pages = {210402},
  numpages = {7},
  year = {2024},
  month = {May},
  publisher = {American Physical Society},
  doi = {10.1103/PhysRevLett.132.210402},
  url = {https://link.aps.org/doi/10.1103/PhysRevLett.132.210402}
}

@article{PhysRevLett.122.210601,
  title = {Dissipative Charging of a Quantum Battery},
  author = {Barra, Felipe},
  journal = {Phys. Rev. Lett.},
  volume = {122},
  issue = {21},
  pages = {210601},
  numpages = {6},
  year = {2019},
  month = {May},
  publisher = {American Physical Society},
  doi = {10.1103/PhysRevLett.122.210601},
  url = {https://link.aps.org/doi/10.1103/PhysRevLett.122.210601}
}

@article{PhysRevResearch.2.013095,
  title = {Stabilizing open quantum batteries by sequential measurements},
  author = {Gherardini, Stefano and Campaioli, Francesco and Caruso, Filippo and Binder, Felix C.},
  journal = {Phys. Rev. Res.},
  volume = {2},
  issue = {1},
  pages = {013095},
  numpages = {9},
  year = {2020},
  month = {Jan},
  publisher = {American Physical Society},
  doi = {10.1103/PhysRevResearch.2.013095},
  url = {https://link.aps.org/doi/10.1103/PhysRevResearch.2.013095}
}

@article{bv4w-jr6q,
  title = {Two-Time Weak-Measurement Protocol for Ergotropy Protection in Open Quantum Batteries},
  author = {Malavazi, Andr\'e H.A. and Sagar, Rishav and Ahmadi, Borhan and Dieguez, Pedro R.},
  journal = {PRX Energy},
  volume = {4},
  issue = {2},
  pages = {023011},
  numpages = {28},
  year = {2025},
  month = {Jun},
  publisher = {American Physical Society},
  doi = {10.1103/bv4w-jr6q},
  url = {https://link.aps.org/doi/10.1103/bv4w-jr6q}
}

@article{Shastri2025,
author={Shastri, Rahul
and Jiang, Chao
and Xu, Guo-Hua
and Prasanna Venkatesh, B.
and Watanabe, Gentaro},
title={Dephasing enabled fast charging of quantum batteries},
journal={npj Quantum Information},
year={2025},
month={Jan},
day={19},
volume={11},
number={1},
pages={9},
issn={2056-6387},
doi={10.1038/s41534-025-00959-5},
url={https://doi.org/10.1038/s41534-025-00959-5}
}

@article{zakavati2025optimizing,
  title={Optimizing the Charging of Open Quantum Batteries using Long Short-Term Memory-Driven Reinforcement Learning},
  author={Zakavati, Shadab and Salimi, Shahriar and Arash, Behrouz},
  journal={arXiv:2504.19840},
  year={2025},
  url={https://doi.org/10.48550/arXiv.2504.19840}
}

@article{PhysRevA.102.052223,
  title = {Environment-mediated charging process of quantum batteries},
  author = {Tabesh, F. T. and Kamin, F. H. and Salimi, S.},
  journal = {Phys. Rev. A},
  volume = {102},
  issue = {5},
  pages = {052223},
  numpages = {8},
  year = {2020},
  month = {Nov},
  publisher = {American Physical Society},
  doi = {10.1103/PhysRevA.102.052223},
  url = {https://link.aps.org/doi/10.1103/PhysRevA.102.052223}
}

@article{PhysRevE.104.064143,
  title = {Enhancing the performance of an open quantum battery via environment engineering},
  author = {Xu, Kai and Zhu, Han-Jie and Zhang, Guo-Feng and Liu, Wu-Ming},
  journal = {Phys. Rev. E},
  volume = {104},
  issue = {6},
  pages = {064143},
  numpages = {9},
  year = {2021},
  month = {Dec},
  publisher = {American Physical Society},
  doi = {10.1103/PhysRevE.104.064143},
  url = {https://link.aps.org/doi/10.1103/PhysRevE.104.064143}
}

@article{PhysRevA.104.032207,
  title = {Fast charging of a quantum battery assisted by noise},
  author = {Ghosh, Srijon and Chanda, Titas and Mal, Shiladitya and Sen(De), Aditi},
  journal = {Phys. Rev. A},
  volume = {104},
  issue = {3},
  pages = {032207},
  numpages = {12},
  year = {2021},
  month = {Sep},
  publisher = {American Physical Society},
  doi = {10.1103/PhysRevA.104.032207},
  url = {https://link.aps.org/doi/10.1103/PhysRevA.104.032207}
}

@article{PhysRevA.105.062203,
  title = {Collective effects and quantum coherence in dissipative charging of quantum batteries},
  author = {Mayo, Franco and Roncaglia, Augusto J.},
  journal = {Phys. Rev. A},
  volume = {105},
  issue = {6},
  pages = {062203},
  numpages = {11},
  year = {2022},
  month = {Jun},
  publisher = {American Physical Society},
  doi = {10.1103/PhysRevA.105.062203},
  url = {https://link.aps.org/doi/10.1103/PhysRevA.105.062203}
}

@article{malavazi2025charge,
  title = {Charge-preserving operations in quantum batteries},
  author = {Malavazi, André H. A. and Ahmadi, Borhan and Horodecki, Paweł and Dieguez, Pedro R.},
  journal = {PRX Energy},
  volume = {5},
  pages = {023004},
  year = {2026},
  month = {Mar},
  publisher = {American Physical Society},
  doi = {10.1103/2jtp-jpkn},
  url = {https://doi.org/10.1103/2jtp-jpkn}
}

@book{Serafini2017QuantumCV,
  title={{Quantum Continuous Variables: A Primer of Theoretical Methods}},
  author={Alessio Serafini},
  publisher={CRC Press},
  year={2017},
  url={https://api.semanticscholar.org/CorpusID:125125509}
}

@article{RevModPhys.96.031001,
  title = {Colloquium: Quantum batteries},
  author = {Campaioli, Francesco and Gherardini, Stefano and Quach, James Q. and Polini, Marco and Andolina, Gian Marcello},
  journal = {Rev. Mod. Phys.},
  volume = {96},
  issue = {3},
  pages = {031001},
  numpages = {30},
  year = {2024},
  month = {Jul},
  publisher = {American Physical Society},
  doi = {10.1103/RevModPhys.96.031001},
  url = {https://link.aps.org/doi/10.1103/RevModPhys.96.031001}
}

@article{PhysRevA.111.042216,
  title = {Efficient wireless charging of a quantum battery},
  author = {Hu, Ming-Liang and Gao, Ting and Fan, Heng},
  journal = {Phys. Rev. A},
  volume = {111},
  issue = {4},
  pages = {042216},
  numpages = {13},
  year = {2025},
  month = {Apr},
  publisher = {American Physical Society},
  doi = {10.1103/PhysRevA.111.042216},
  url = {https://link.aps.org/doi/10.1103/PhysRevA.111.042216}
}

@article{PhysRevLett.102.207209,
  title = {Quantum Noise Interference and Backaction Cooling in Cavity Nanomechanics},
  author = {Elste, Florian and Girvin, S. M. and Clerk, A. A.},
  journal = {Phys. Rev. Lett.},
  volume = {102},
  issue = {20},
  pages = {207209},
  numpages = {4},
  year = {2009},
  month = {May},
  publisher = {American Physical Society},
  doi = {10.1103/PhysRevLett.102.207209},
  url = {https://link.aps.org/doi/10.1103/PhysRevLett.102.207209}
}

@article{PhysRevLett.121.137203,
  title = {Level Attraction Due to Dissipative Magnon-Photon Coupling},
  author = {Harder, M. and Yang, Y. and Yao, B. M. and Yu, C. H. and Rao, J. W. and Gui, Y. S. and Stamps, R. L. and Hu, C.-M.},
  journal = {Phys. Rev. Lett.},
  volume = {121},
  issue = {13},
  pages = {137203},
  numpages = {6},
  year = {2018},
  month = {Sep},
  publisher = {American Physical Society},
  doi = {10.1103/PhysRevLett.121.137203},
  url = {https://link.aps.org/doi/10.1103/PhysRevLett.121.137203}
}

@article{PhysRevLett.126.163604,
  title = {Dissipatively Controlled Optomechanical Interaction via Cascaded Photon-Phonon Coupling},
  author = {Shen, Zhen and Zhang, Yan-Lei and Zou, Chang-Ling and Guo, Guang-Can and Dong, Chun-Hua},
  journal = {Phys. Rev. Lett.},
  volume = {126},
  issue = {16},
  pages = {163604},
  numpages = {6},
  year = {2021},
  month = {Apr},
  publisher = {American Physical Society},
  doi = {10.1103/PhysRevLett.126.163604},
  url = {https://link.aps.org/doi/10.1103/PhysRevLett.126.163604}
}

@article{Zhang2022PhononLasing,
  title = {Dissipative coupling-induced phonon lasing},
  author = {Zhang, Qiankun and Yang, Cheng and Sheng, Jiteng and Wu, Haibin},
  journal = {Proc. Natl. Acad. Sci. U.S.A.},
  volume = {119},
  number = {52},
  pages = {e2207543119},
  year = {2022},
  doi = {10.1073/pnas.2207543119},
  url = {https://doi.org/10.1073/pnas.2207543119}
}

@article{Wang2025MagnonPolaritonLasing,
  title = {Single-Mode Magnon-Polariton Lasing and Amplification Controlled by Dissipative Coupling},
  author = {Wang, Zi-Qi and Wang, Zi-Yuan and Wang, Yi-Pu and You, J. Q.},
  journal = {Phys. Rev. Lett.},
  volume = {135},
  issue = {18},
  pages = {186704},
  year = {2025},
  month = {Oct},
  publisher = {American Physical Society},
  doi = {10.1103/bnyn-mbwv},
  url = {https://link.aps.org/doi/10.1103/bnyn-mbwv}
}

@article{Guo2024LevelPinningWPT,
  title = {Level pinning of anti-{PT}-symmetric circuits for efficient wireless power transfer},
  author = {Guo, Zhiwei and Yang, Fengqing and Zhang, Haiyan and Wu, Xian and Wu, Qiong and Zhu, Kejia and Jiang, Jun and Jiang, Haitao and Yang, Yaping and Li, Yunhui and Chen, Hong},
  journal = {National Science Review},
  volume = {11},
  number = {1},
  pages = {nwad172},
  year = {2024},
  doi = {10.1093/nsr/nwad172},
  url = {https://doi.org/10.1093/nsr/nwad172}
}

@article{Hymas2026SuperextensiveQB,
  title = {Superextensive electrical power from a quantum battery},
  author = {Hymas, Kieran and Muir, Jack B. and Tibben, Daniel and van Embden, Joel and Hirai, Tadahiko and Dunn, Christopher J. and G{\'o}mez, Daniel E. and Hutchison, James A. and Smith, Trevor A. and Quach, James Q.},
  journal = {Light: Science \& Applications},
  volume = {15},
  pages = {168},
  year = {2026},
  doi = {10.1038/s41377-026-02240-6},
  url = {https://doi.org/10.1038/s41377-026-02240-6}
}

\clearpage
\onecolumngrid  


\setcounter{section}{0}
\setcounter{subsection}{0}
\setcounter{equation}{0}
\setcounter{figure}{0}
\setcounter{table}{0}

\renewcommand{\theequation}{S\arabic{equation}}
\renewcommand{\thefigure}{\arabic{figure}}
\renewcommand{\thetable}{S\arabic{table}}

\newcommand{\suppnote}[1]{%
    \refstepcounter{section}%
    \setcounter{subsection}{0}%
    \section*{Supplementary Note \arabic{section}. #1}%
    \addcontentsline{toc}{section}{Supplementary Note \arabic{section}. #1}%
}

\newcommand{\suppsubsection}[1]{%
    \refstepcounter{subsection}%
    \subsection*{\Alph{subsection}. #1}%
    \addcontentsline{toc}{subsection}{\Alph{subsection}. #1}%
}

\suppnote{Parametric-pump instability threshold for three benchmark models}
\label{sec:parametric_pump_threshold_three_cases}

\subsection{General Hamiltonian and master equation}
\label{sec:general_model_pump_dissipation}

We consider bosonic modes \(\hat a\), \(\hat b\), and, when present, an auxiliary mediator mode \(\hat c\), satisfying the canonical commutation relations
\begin{equation}
[\hat a,\hat a^\dagger]=[\hat b,\hat b^\dagger]=[\hat c,\hat c^\dagger]=1,
\qquad
[\hat a,\hat b]=[\hat a,\hat c]=[\hat b,\hat c]=0.
\end{equation}
The dynamics is generated by a quadratic Hamiltonian together with linear Lindblad dissipators. The Hamiltonian is written as
\begin{equation}
\hat H(t) = \hat H_0 +\hat H_c+ \hat H_{\rm int} + \hat H_{\rm pump} + \hat H_{\rm seed}(t),
\label{eq:H_general_split}
\end{equation}
where
\begin{equation}
\hat H_0
=
\delta_a \hat a^\dagger \hat a
+
\delta_b \hat b^\dagger \hat b, \qquad\qquad
\hat H_c =
\delta_c \hat c^\dagger \hat c
\label{eq:H0_general}
\end{equation}
contains the bare mode energies (or detunings),
\begin{equation}
\hat H_{\rm pump}
=
\mathcal E\,\hat a^{\dagger 2}
+
\mathcal E^\ast \hat a^2
\label{eq:Hpump_general}
\end{equation}
is the single-mode two-photon pump acting on mode \(\hat a\). The short seed pulse is taken in the form
\begin{equation}
\hat H_{\rm seed}(t)
=
\varepsilon(t)\,(\hat a+\hat a^\dagger),
\qquad
\varepsilon(t)
=
\varepsilon_0
\exp\!\left[
-\frac{(t-t_0)^2}{2\sigma^2}
\right].
\label{eq:SM_Hseed}
\end{equation}
The reason for having a seed, as shown below, is that a quadratic pump does not displace the state in phase space, but it deforms the fluctuations. In phase-space language, the Wigner function gets stretched, but its center stays at zero. So the center can stay at zero while one quadrature variance decreases and the other increases. Consequently, in the absence of the seed pulse and starting from vacuum, all first moments remain zero, while the second moments may still grow strongly through squeezing and noise amplification. This is the reason why a nonzero coherent contribution, measured for example by \(|\langle \hat b(t)\rangle|^2\), requires either a nonzero initial displacement or an external short linear seed such as \eqref{eq:SM_Hseed}. The interaction Hamiltonian \(\hat H_{\rm int}\) specifies the following coupling architectures:

\begin{description}
    
\item[Case I] For the two-mode coherent benchmark, the beam-splitter interaction Hamiltonian is
\begin{equation}
\hat H_{\rm int}^{\rm coh} = J \hat a^\dagger \hat b + J^\ast \hat a \hat b^\dagger,
\label{eq:Hint_coh_general}
\end{equation}
with \(J\in\mathbb C\). In this case, the auxiliary mode \(\hat c\) is absent and one may simply set \(\delta_c=0\);

\item[Case II] For the three-mode dissipative engineered model, there is no direct coherent coupling between \(\hat a\) and \(\hat b\). Instead, the modes are effectively coupled through dissipative channels mediated by \(\hat c\). Thus, in this case, we may take
\begin{equation}
\hat H_{\rm int}^{3\rm diss}=0,
\label{eq:Hint_3mode_diss_general}
\end{equation}
and all mode evolution is generated by the dissipators;

\item[Case III] For the reduced two-mode dissipative model obtained after eliminating \(\hat c\), one also does not have a coherent beam-splitter term in the effective Hamiltonian, so
\begin{equation}
\hat H_{\rm int}^{{\rm red}\,{\rm diss}}=0,
\label{eq:Hint_reduced_diss_general}
\end{equation}
while the effective dissipative coupling appears directly in the reduced drift matrix.
\end{description}
Apart from different interaction terms, the considered cases differ through dissipative terms in the dynamics governed by the Markovian master equation of GKLS form
\begin{equation}
\dot{\hat{\rho}} = -i[\hat H(t),\hat \rho] + \sum_\ell \mathcal D[\hat L_\ell]\hat \rho,
\label{eq:ME_general}
\end{equation}
with dissipators defined through
\begin{equation}
\mathcal D[\hat L]\hat\rho = \hat L \hat\rho \hat L^\dagger
- \frac12\{\hat L^\dagger \hat L,\hat\rho\}.
\label{eq:Dissipator_general}
\end{equation}
The local dissipative channels are taken to be
\begin{equation}
\hat L_a^{\rm loss}=\sqrt{\kappa_a}\,\hat a,
\quad
\hat L_b^{\rm loss}=\sqrt{\kappa_b}\,\hat b,
\quad
\hat L_c^{\rm loss}=\sqrt{\kappa_c}\,\hat c.
\label{eq:local_losses_general}
\end{equation}

\begin{description}
\item[Case I] For the two-mode coherent benchmark, one accounts for the interaction Hamiltonian $\hat H_{\rm int}^{\rm coh}$ to arrive at
\begin{align}
\dot{\hat \rho}
&=
-i[\hat H_0+\hat H_{\rm int}^{\rm coh}+\hat H_{\rm pump}+\hat H_{\rm seed}(t),\hat \rho]
+\kappa_a \mathcal D[\hat a]\hat \rho
+\kappa_b \mathcal D[\hat b]\hat \rho,
\label{eq:ME_2mode_coh_full}
\end{align}
i.e., with no shared dissipator and no mediator mode.

\item[Case II]
For the three-mode dissipative engineered model, the additional shared dissipative channels are considered:
\begin{equation}
\hat L_{z_a}=\sqrt{\Gamma}\,\hat z_a,
\quad
\hat L_{z_b}=\sqrt{\Gamma}\,\hat z_b,
\label{eq:shared_L_general}
\end{equation}
with operators
\begin{equation}
\hat z_a = p_a \hat a + p_{ca}\hat c,
\quad
\hat z_b = p_b \hat b + p_{cb}\hat c.
\label{eq:shared_jumps_general}
\end{equation}
The master equation then becomes
\begin{align}
\dot{\hat \rho} &= -i[\hat H_0+\hat H_c+\hat H_{\rm pump}+\hat H_{\rm seed}(t),\hat \rho]
+\kappa_a \mathcal D[\hat a]\hat \rho
+\kappa_b \mathcal D[\hat b]\hat \rho
+\kappa_c \mathcal D[\hat c]\hat \rho
+\Gamma \mathcal D[\hat z_a]\hat \rho
+\Gamma \mathcal D[\hat z_b]\hat \rho.
\label{eq:ME_3mode_diss_full}
\end{align}

\item[Case III]
Finally, after eliminating the auxiliary mode \(\hat c\) from Eq. (\ref{eq:ME_3mode_diss_full}), we obtain
\begin{equation}
\dot\rho_{ab} = -i[\hat H_{\rm red}(t),\rho_{ab}] + \sum_{i,j=a,b}
(\mathsf K_{\rm red})_{ij}\left(\hat d_i \rho_{ab} \hat d_j^\dagger
-\frac12\{\hat d_j^\dagger \hat d_i,\rho_{ab}\}\right),
\end{equation}
where
\begin{equation}
\mathsf K_{\rm red}
=
\begin{pmatrix}
2\gamma_a & -2g\\
-2g^\ast & 2\gamma_b
\end{pmatrix},
\qquad
\hat{\mathbf d} =
\begin{pmatrix}
\hat a\\
\hat b
\end{pmatrix},
\qquad
\hat H_{\rm red}(t) = \delta\,\hat a^\dagger \hat a - \delta\,\hat b^\dagger \hat b + \hat H_{\rm pump}+\hat H_{\rm seed}(t).
\end{equation}
\end{description}

Based on the above equations, one may derive equations of motion for the first and second moments. Since all Hamiltonian and Lindblad operators are at most quadratic or linear in the bosonic operators, the moment hierarchy closes exactly at second order, and the dynamics is Gaussian~\cite{Serafini2017QuantumCV}.

\suppnote{Closed equations of motion for first and second moments}
\label{sec:SM_moment_equations}

In this section, we collect the closed equations of motion for the first and second moments for the three models studied in the main text. 
Throughout, we use the standard Lindblad convention
\begin{equation}
\mathcal D[\hat L]\rho
=
\hat L\rho \hat L^\dagger
-\frac12\{\hat L^\dagger \hat L,\rho\},
\label{eq:SM_Dissipator}
\end{equation}
and the single-mode two-photon pump
\begin{equation}
\hat H_{\rm pump}
=
\mathcal E\,\hat a^{\dagger 2}
+
\mathcal E^\ast \hat a^2.
\label{eq:SM_Hpump}
\end{equation}
The short seed pulse is taken in the form
\begin{equation}
\hat H_{\rm seed}(t)
=
\varepsilon(t)\,(\hat a+\hat a^\dagger),
\qquad
\varepsilon(t)
=
\varepsilon_0
\exp\!\left[
-\frac{(t-t_0)^2}{2\sigma^2}
\right].
\label{eq:SM_Hseed2}
\end{equation}
The total Hamiltonian in each case is the sum of the model-specific quadratic Hamiltonian, the pump term \eqref{eq:SM_Hpump}, and the seed term \eqref{eq:SM_Hseed2}.

For a Gaussian system with a quadratic Hamiltonian and linear Lindblad operators, the first moments and the second moments form a closed hierarchy. We define the first-moment vector by
\begin{equation}
m(t)=
\begin{pmatrix}
\langle \hat a\rangle \\
\langle \hat b\rangle \\
\langle \hat c\rangle
\end{pmatrix},
\end{equation}
with the obvious truncation to the relevant modes in the two-mode cases. For the second moments, we use the normally ordered set
\begin{equation}
n_i:=\langle \hat i^\dagger \hat i\rangle,
\qquad
s_{ij}:=\langle \hat i^\dagger \hat j\rangle,
\qquad
u_i:=\langle \hat i^2\rangle,
\qquad
u_{ij}:=\langle \hat i \hat j\rangle,
\label{eq:SM_second_moments_defs}
\end{equation}
with \(i,j\in\{a,b,c\}\) as appropriate. The equations for the conjugate moments follow by complex conjugation.
The seed pulse contributes only to the first moments. Indeed,
\begin{equation}
\frac{d}{dt}\langle \hat O\rangle_{\rm seed}
=
i\langle[\hat H_{\rm seed}(t),\hat O]\rangle,
\end{equation}
and using $[\hat a+\hat a^\dagger,\hat a]=-1$ and $[\hat a+\hat a^\dagger,\hat a^\dagger]=1$, one finds
\begin{equation}
\frac{d}{dt}\langle \hat a\rangle_{\rm seed}
=
-i\,\varepsilon(t),
\qquad
\frac{d}{dt}\langle \hat a^\dagger\rangle_{\rm seed}
=
+i\,\varepsilon(t),
\label{eq:SM_seed_firstmom}
\end{equation}
while
\begin{equation}
\frac{d}{dt}\langle \hat b\rangle_{\rm seed}
=
\frac{d}{dt}\langle \hat c\rangle_{\rm seed}
=
\frac{d}{dt}\langle \hat b^\dagger\rangle_{\rm seed}
=
\frac{d}{dt}\langle \hat c^\dagger\rangle_{\rm seed}
=
0.
\label{eq:SM_seed_rest}
\end{equation}
By contrast, the two-photon pump \eqref{eq:SM_Hpump} is purely quadratic. Indeed, as mentioned above, it modifies the homogeneous drift matrix of the first moments but does not generate any inhomogeneous driving term. In other words, a quadratic pump does not displace the state in phase space, but it deforms the fluctuations. 
%
%
%
%

%
\subsection{Case I: Two-mode coherent interaction}
\label{sec:SM_case_coherent}

The master equation is
\begin{equation}
\dot\rho
=
-i[\hat H_{\rm coh}(t),\rho]
+
\kappa_a \mathcal D[\hat a]\rho
+
\kappa_b \mathcal D[\hat b]\rho,
\label{eq:SM_ME_coh}
\end{equation}
with
\begin{equation}
\hat{H}_{{\rm coh}}(t)=\delta_{a}\hat{a}^{\dagger}\hat{a}+\delta_{b}\hat{b}^{\dagger}\hat{b}+J\hat{a}^{\dagger}\hat{b}+J^{\ast}\hat{a}\hat{b}^{\dagger}+\mathcal{E}\hat{a}^{\dagger2}+\mathcal{E}^{\ast}\hat{a}^{2}+\varepsilon(t)(\hat{a}+\hat{a}^{\dagger}).
\label{eq:SM_H_coh}
\end{equation}
%
%


\subsubsection{First moments}

Let us define $\gamma_{a,b}=\kappa_{a,b}/2$. Then, the first moments satisfy
\begin{align}
\frac{d}{dt}\langle\hat{a}\rangle&=-\left(\gamma_{a}+i\delta_{a}\right)\langle\hat{a}\rangle-iJ\langle\hat{b}\rangle-2i\mathcal{E}\langle\hat{a}^{\dagger}\rangle-i\varepsilon(t),&\frac{d}{dt}\langle\hat{b}\rangle&=-\left(\gamma_{b}+i\delta_{b}\right)\langle\hat{b}\rangle-iJ^{\ast}\langle\hat{a}\rangle,\\
\frac{d}{dt}\langle\hat{a}^{\dag}\rangle&=-\left(\gamma_{a}-i\delta_{a}\right)\langle\hat{a}^{\dag}\rangle+iJ^{*}\langle\hat{b}^{\dag}\rangle+2i\mathcal{E}^{*}\langle\hat{a}\rangle+i\varepsilon(t),&
\frac{d}{dt}\langle\hat{b}^{\dag}\rangle&=-\left(\gamma_{b}-i\delta_{b}\right)\langle\hat{b}^{\dag}\rangle+iJ\langle\hat{a}^{\dag}\rangle.
\label{eq:SM_coh_bd}
\end{align}

\subsubsection{Second moments}

Introducing
\begin{equation}
n_a=\langle \hat a^\dagger \hat a\rangle,
\qquad
n_b=\langle \hat b^\dagger \hat b\rangle,
\qquad
s_{ab}=\langle \hat a^\dagger \hat b\rangle,
\qquad
u_a=\langle \hat a^2\rangle,
\qquad
u_b=\langle \hat b^2\rangle,
\qquad
u_{ab}=\langle \hat a \hat b\rangle,
\label{eq:SM_coh_second_defs}
\end{equation}
we have
\begin{align}
\dot{n_{a}}&=-2\gamma_{a}n_{a}-iJs_{ab}+iJ^{\ast}s^{*}_{ab}-2i\mathcal{E}u^{*}_{a}+2i\mathcal{E}^{\ast}u_{a}+i\varepsilon(t)\langle\hat{a}\rangle-i\varepsilon(t)\langle\hat{a}^{\dagger}\rangle,\nonumber\\
\dot{n_{b}}&=-2\gamma_{b}n_{b}+iJs_{ab}-iJ^{\ast}s^{*}_{ab},\nonumber
\\\dot{s_{ab}}&=\left[-(\gamma_{a}+\gamma_{b})+i(\delta_{a}-\delta_{b})\right]s_{ab}+iJ^{\ast}(n_{b}-n_{a})+2i\mathcal{E}^{\ast}u_{ab}+i\varepsilon(t)\langle\hat{b}\rangle,\nonumber\\
\dot{u_{a}}&=-2(\gamma_{a}+i\delta_{a})u_{a}-2iJu_{ab}-2i\mathcal{E}(2n_{a}+1)-2i\varepsilon(t)\langle\hat{a}\rangle,\\\dot{u_{b}}&=-2(\gamma_{b}+i\delta_{b})u_{b}-2iJ^{\ast}u_{ab},\nonumber\\
\dot{u_{ab}}&=\left[-(\gamma_{a}+\gamma_{b})-i(\delta_{a}+\delta_{b})\right]u_{ab}-iJu_{b}-iJ^{\ast}u_{a}-2i\mathcal{E}s_{ab}-i\varepsilon(t)\langle\hat{b}\rangle.\nonumber
\label{eq:SM_coh_uab}
\end{align}
The dynamical equations for \(s_{ab}^\ast\), \(u_a^\ast\), \(u_b^\ast\), and \(u_{ab}^\ast\) follow by complex conjugation.


\subsection{Case II: Three-mode dissipative engineered interaction}
\label{sec:SM_case_3mode}

The master equation is
\begin{equation}
\dot\rho
=
-i[\hat H_{3{\rm m}}(t),\rho]
+
\kappa_a \mathcal D[\hat a]\rho
+
\kappa_b \mathcal D[\hat b]\rho
+
\kappa_c \mathcal D[\hat c]\rho
+
\Gamma \mathcal D[\hat z_a]\rho
+
\Gamma \mathcal D[\hat z_b]\rho,
\label{eq:SM_ME_3mode}
\end{equation}
with
\begin{equation}
    \hat{H}_{3{\rm m}}(t)=\delta_{a}\hat{a}^{\dagger}\hat{a}+\delta_{b}\hat{b}^{\dagger}\hat{b}+\delta_{c}\hat{c}^{\dagger}\hat{c}+\mathcal{E}\hat{a}^{\dagger2}+\mathcal{E}^{\ast}\hat{a}^{2}+\varepsilon(t)(\hat{a}+\hat{a}^{\dagger}),
\end{equation}
and
\begin{equation}
\hat z_a=p_a \hat a+p_{ca}\hat c,
\qquad
\hat z_b=p_b \hat b+p_{cb}\hat c.
\label{eq:SM_za_zb}
\end{equation}

In the standard Lindblad convention, the effective first-moment decay rates and dissipative couplings are
\begin{equation}
\gamma_a=\frac{\kappa_a}{2}+\frac{\Gamma}{2}|p_a|^2,
\qquad
\gamma_b=\frac{\kappa_b}{2}+\frac{\Gamma}{2}|p_b|^2,
\qquad
\gamma_c=\frac{\kappa_c}{2}+\frac{\Gamma}{2}\bigl(|p_{ca}|^2+|p_{cb}|^2\bigr),
\label{eq:SM_gammas_3mode}
\end{equation}
\begin{equation}
g_{ac}=\frac{\Gamma}{2}p_a^\ast p_{ca},
\qquad
g_{bc}=\frac{\Gamma}{2}p_b^\ast p_{cb}.
\label{eq:SM_gs_3mode}
\end{equation}

\subsubsection{First moments}

The first moments satisfy
\begin{equation}
\begin{aligned}
    \frac{d}{dt}\langle\hat{a}\rangle&=-\left(\gamma_{a}+i\delta_{a}\right)\langle\hat{a}\rangle-g_{ac}\langle\hat{c}\rangle-2i\mathcal{E}\langle\hat{a}^{\dagger}\rangle-i\varepsilon(t),&\frac{d}{dt}\langle\hat{a}^{\dag}\rangle&=-\left(\gamma_{a}-i\delta_{a}\right)\langle\hat{a}^{\dag}\rangle-g^{*}_{ac}\langle\hat{c}^{\dag}\rangle+2i\mathcal{E}^{*}\langle\hat{a}\rangle+i\varepsilon(t),\\\frac{d}{dt}\langle\hat{b}\rangle&=-\left(\gamma_{b}+i\delta_{b}\right)\langle\hat{b}\rangle-g_{bc}\langle\hat{c}\rangle,&\frac{d}{dt}\langle\hat{b}^{\dag}\rangle&=-\left(\gamma_{b}-i\delta_{b}\right)\langle\hat{b}^{\dag}\rangle-g^{*}_{bc}\langle\hat{c}^{\dag}\rangle,\\\frac{d}{dt}\langle\hat{c}\rangle&=-\left(\gamma_{c}+i\delta_{c}\right)\langle\hat{c}\rangle-g^{*}_{ac}\langle\hat{a}\rangle-g^{*}_{bc}\langle\hat{b}\rangle,&\frac{d}{dt}\langle\hat{c}^{\dag}\rangle&=-\left(\gamma_{c}-i\delta_{c}\right)\langle\hat{c}^{\dag}\rangle-g_{ac}\langle\hat{a}^{\dag}\rangle-g_{bc}\langle\hat{b}^{\dag}\rangle.
\end{aligned}
\end{equation}

\subsubsection{Second moments}

Similarly, by introducing
\begin{equation}
\begin{aligned}  n_{a}&=\langle\hat{a}^{\dagger}\hat{a}\rangle,&n_{b}&=\langle\hat{b}^{\dagger}\hat{b}\rangle,&n_{c}&=\langle\hat{c}^{\dagger}\hat{c}\rangle,&s_{ab}&=\langle\hat{a}^{\dagger}\hat{b}\rangle,&s_{ac}&=\langle\hat{a}^{\dagger}\hat{c}\rangle,&s_{bc}&=\langle\hat{b}^{\dagger}\hat{c}\rangle,\\u_{a}&=\langle\hat{a}^{2}\rangle,&u_{b}&=\langle\hat{b}^{2}\rangle,&u_{c}&=\langle\hat{c}^{2}\rangle,&u_{ab}&=\langle\hat{a}\hat{b}\rangle,&u_{ac}&=\langle\hat{a}\hat{c}\rangle,&u_{bc}&=\langle\hat{b}\hat{c}\rangle.
\end{aligned}
\end{equation}
One can show that
\begin{align*}
    \dot{n_{a}}&=-2\gamma_{a}n_{a}-g_{ac}s_{ac}-g^{*}_{ac}s^{*}_{ac}-2i\mathcal{E}u^{*}_{a}+2i\mathcal{E}^{\ast}u_{a}-i\varepsilon(t)\langle\hat{a}^{\dagger}\rangle+i\varepsilon(t)\langle\hat{a}\rangle,\\\dot{n_{b}}&=-2\gamma_{b}n_{b}-g_{bc}s_{bc}-g^{*}_{bc}s^{*}_{bc},\\\dot{n_{c}}&=-2\gamma_{c}n_{c}-g_{ac}s_{ac}-g^{*}_{ac}s^{*}_{ac}-g_{bc}s_{bc}-g^{*}_{bc}s^{*}_{bc},\\\dot{s_{ab}}&=\left[-\left(\gamma_{a}+\gamma_{b}\right)+i\left(\delta_{a}-\delta_{b}\right)\right]s_{ab}-g^{*}_{ac}s^{*}_{bc}-g_{bc}s_{ac}+2i\mathcal{E}^{\ast}u_{ab}+i\varepsilon(t)\langle\hat{b}\rangle,\\\dot{s_{ac}}&=\left[-\left(\gamma_{a}+\gamma_{c}\right)+i\left(\delta_{a}-\delta_{c}\right)\right]s_{ac}-g^{*}_{ac}\left(n_{a}+n_{c}\right)-g^{*}_{bc}s_{ab}+2i\mathcal{E}^{\ast}u_{ac}+i\varepsilon(t)\langle\hat{c}\rangle,\\\dot{s_{bc}}&=\left[-\left(\gamma_{b}+\gamma_{c}\right)+i\left(\delta_{b}-\delta_{c}\right)\right]s_{bc}-g^{*}_{bc}\left(n_{b}+n_{c}\right)-g^{*}_{ac}s^{*}_{ab},\\\dot{u_{a}}&=-2\left(\gamma_{a}+i\delta_{a}\right)u_{a}-g_{ac}u_{ac}-2i\mathcal{E}\left(2n_{a}+1\right)-2i\varepsilon(t)\langle\hat{a}\rangle,\numberthis\\
    \dot{u_{b}}&=-2\left(\gamma_{b}+i\delta_{b}\right)u_{b}-2g_{bc}u_{bc},\\ \dot{u_{c}}&=-2\left(\gamma_{c}+i\delta_{c}\right)u_{c}-2g^{*}_{ac}u_{ac}-2g^{*}_{bc}u_{bc},\\\dot{u_{ab}}&=\left[-\left(\gamma_{a}+\gamma_{b}\right)-i\left(\delta_{a}+\delta_{b}\right)\right]u_{ab}-g_{ac}u_{bc}-g_{bc}u_{ac}-2i\mathcal{E}s_{ab}-i\varepsilon(t)\langle\hat{b}\rangle,\\\dot{u_{ac}}&=\left[-\left(\gamma_{a}+\gamma_{c}\right)-i\left(\delta_{a}+\delta_{c}\right)\right]u_{ac}-g_{ac}u_{c}-g^{*}_{ac}u_{a}-g^{*}_{bc}u_{ab}-2i\mathcal{E}s_{ac}-i\varepsilon(t)\langle\hat{c}\rangle,\\\dot{u_{bc}}&=\left[-\left(\gamma_{b}+\gamma_{c}\right)-i\left(\delta_{b}+\delta_{c}\right)\right]u_{bc}-g_{bc}u_{c}-g^{*}_{bc}u_{b}-g^{*}_{ac}u_{ab}.
\end{align*}
As before, the equations for the conjugate moments follow by complex conjugation.


\subsection{Case III: Reduced two-mode dissipative model}
\label{sec:SM_case_reduced}

We now derive the reduced Gaussian generator for the two-mode dissipative model obtained by eliminating the auxiliary mode \(\hat c\). The starting point is the three-mode master equation
\begin{equation}
\dot\rho =
-i[\hat H_{3{\rm m}}(t),\rho]
+
\kappa_a \mathcal D[\hat a]\rho
+
\kappa_b \mathcal D[\hat b]\rho
+
\kappa_c \mathcal D[\hat c]\rho
+
\Gamma \mathcal D[\hat z_a]\rho
+
\Gamma \mathcal D[\hat z_b]\rho,
\label{eq:SM_ME_full3_again}
\end{equation}
with
\begin{equation}
\hat z_a=p_a \hat a+p_{ca}\hat c,
\qquad
\hat z_b=p_b \hat b+p_{cb}\hat c,
\label{eq:SM_za_zb_again}
\end{equation}
and
\begin{equation}
\hat H_{3{\rm m}}(t) =
\delta_a\,\hat a^\dagger \hat a
+\delta_b\,\hat b^\dagger \hat b
+ \delta_c\,\hat c^\dagger \hat c
+ \mathcal E\,\hat a^{\dagger 2}
+
\mathcal E^\ast \hat a^2
+
\varepsilon(t)(\hat a+\hat a^\dagger).
\label{eq:SM_H3_again}
\end{equation}

Let
\begin{equation}
\hat{\mathbf d} =
\begin{pmatrix}
\hat a\\
\hat b\\
\hat c
\end{pmatrix}.
\end{equation}
and, for simplicity, let us also assume $\delta_{c} = 0$.
The dissipative part of Eq.~\eqref{eq:SM_ME_full3_again} can be written in Gaussian Kossakowski form,
\begin{equation}
\mathcal L_{\rm diss}\rho =
\sum_{i,j=a,b,c}
\mathsf K_{ij}
\left(
\hat d_i \rho \hat d_j^\dagger
-\frac12\{\hat d_j^\dagger \hat d_i,\rho\}
\right),
\label{eq:SM_Kossakowski_full}
\end{equation}
where the positive semidefinite Kossakowski matrix is
\begin{equation}
\mathsf K =
\begin{pmatrix}
\kappa_a+\Gamma_a & 0 & \Gamma \mu_{ca}\\
0 & \kappa_b+\Gamma_b & \Gamma \mu_{cb}\\
\Gamma \mu_{ca}^\ast & \Gamma \mu_{cb}^\ast & \kappa_c+\Gamma_{ca}+\Gamma_{cb}
\end{pmatrix},
\label{eq:SM_K_full}
\end{equation}
with \(\Gamma_a=\Gamma |p_a|^2\), \(\Gamma_b=\Gamma |p_b|^2\), \(\Gamma_{ca}=\Gamma |p_{ca}|^2\), \(\Gamma_{cb}=\Gamma |p_{cb}|^2\), \(\mu_{ca}=p_{ca}p_a^\ast\), and \(\mu_{cb}=p_{cb}p_b^\ast\). We partition \(\mathsf K\) into the slow sector \(a,b\) and the fast sector \(c\),
\begin{equation}
\mathsf K =
\begin{pmatrix}
\mathsf K_{\rm ss} & \mathsf K_{\rm sf}\\
\mathsf K_{\rm fs} & \mathsf K_{\rm ff}
\end{pmatrix},
\label{eq:SM_K_block}
\end{equation}
where
\begin{equation}
\mathsf K_{\rm ss} =
\begin{pmatrix}
\kappa_a+\Gamma_a & 0\\
0 & \kappa_b+\Gamma_b
\end{pmatrix},
\qquad
\mathsf K_{\rm sf}=
\begin{pmatrix}
\Gamma \mu_{ca}\\
\Gamma \mu_{cb}
\end{pmatrix},
\qquad
\mathsf K_{\rm fs}=\mathsf K_{\rm sf}^\dagger,
\qquad
\mathsf K_{\rm ff}=\kappa_c+\Gamma_{ca}+\Gamma_{cb}.
\label{eq:SM_K_blocks_explicit}
\end{equation}

We now derive the reduced dissipative kernel. In the passive part of the annihilation-sector first-moment equation, the dissipative drift matrix generated by Eq.~\eqref{eq:SM_Kossakowski_full} is \(-\mathsf K/2\). Thus, before including the two-photon pump in the slow-sector drift, the dissipative contribution generated by Eq.~\eqref{eq:SM_Kossakowski_full} gives
\begin{equation}
\dot m_s
=
-\frac{1}{2}K_{\rm ss}m_s
-\frac{1}{2}K_{\rm sf}m_f
+
F_s(t),
\qquad
\dot m_f
=
-\frac{1}{2}K_{\rm ff}m_f
-\frac{1}{2}K_{\rm fs}m_s .
\label{eq:SM_slow_fast_first_moments}
\end{equation}
Here \(m_s=(\langle\hat a\rangle,\langle\hat b\rangle)^T\), \(m_f=\langle\hat c\rangle\), and \(F_s(t)\) collects the slow-sector terms not involved in the dissipative elimination, including the detuning Hamiltonian, the weak seed, and, when retained at this stage, the two-photon pairing term coupling \(m_s\) to \(m_s^\ast\). There is no corresponding \(F_f(t)\) term in the fast equation because we take \(\delta_c=0\), the seed acts only on \(\hat a\), and the two-photon pump acts only on \(\hat a\). Thus the fast mediator is driven only through its dissipative coupling to the slow modes.

In the fast-mediator regime, \(K_{\rm ff}\) is the largest relaxation scale, so \(m_f\) follows the slow variables algebraically at leading Markov order. Setting \(\dot m_f\simeq0\) in Eq.~\eqref{eq:SM_slow_fast_first_moments} gives
\begin{equation}
0
=
-\frac{1}{2}K_{\rm ff}m_f
-\frac{1}{2}K_{\rm fs}m_s,
\qquad
m_f
=
-K_{\rm ff}^{-1}K_{\rm fs}m_s .
\label{eq:SM_fast_solution}
\end{equation}
Substituting Eq.~\eqref{eq:SM_fast_solution} into the slow equation gives
\begin{equation}
\dot m_{\rm s}
=
-\frac12
\left(
\mathsf K_{\rm ss}
-
\mathsf K_{\rm sf}\mathsf K_{\rm ff}^{-1}\mathsf K_{\rm fs}
\right)m_{\rm s}
+ F_s(t).
\label{eq:SM_reduced_slow_drift}
\end{equation}
Hence, the effective reduced dissipative kernel is the Schur complement of the fast block \(\mathsf K_{\rm ff}\) in the full Kossakowski matrix \(\mathsf K\) \cite{Roger1985},
\begin{equation}
\mathsf K_{\rm red}
=
\mathsf K_{\rm ss}
-
\mathsf K_{\rm sf}\mathsf K_{\rm ff}^{-1}\mathsf K_{\rm fs}.
\label{eq:SM_Kred_schur}
\end{equation}
Because \(\mathsf K\ge 0\) and \(\mathsf K_{\rm ff}>0\), this Schur complement is also positive semidefinite. This guarantees that the reduced dissipative generator remains completely positive.
Evaluating Eq.~\eqref{eq:SM_Kred_schur} gives
\begin{equation}
\mathsf K_{\rm red}
=
\begin{pmatrix}
\kappa_a+\Gamma_a-\Gamma_{\rm eff}|\mu_{ca}|^2
&
-\Gamma_{\rm eff}\mu_{ca}\mu_{cb}^\ast
\\
-\Gamma_{\rm eff}\mu_{ca}^\ast\mu_{cb}
&
\kappa_b+\Gamma_b-\Gamma_{\rm eff}|\mu_{cb}|^2
\end{pmatrix},
\label{eq:SM_Kred_explicit}
\end{equation}
where
\begin{equation}
\Gamma_{\rm eff}
=
\frac{\Gamma^2}{\kappa_c+\Gamma_{ca}+\Gamma_{cb}}.
\label{eq:SM_Gamma_eff}
\end{equation}
With
\begin{equation}
2\gamma_a
=
\kappa_a+\Gamma_a-\Gamma_{\rm eff}|\mu_{ca}|^2,
\qquad
2\gamma_b
=
\kappa_b+\Gamma_b-\Gamma_{\rm eff}|\mu_{cb}|^2,
\qquad
g
=
\frac{\Gamma_{\rm eff}}{2}\mu_{ca}\mu_{cb}^\ast,
\label{eq:SM_gamma_g_def}
\end{equation}
the reduced kernel becomes
\begin{equation}
\mathsf K_{\rm red}
=
\begin{pmatrix}
2\gamma_a & -2g\\
-2g^\ast & 2\gamma_b
\end{pmatrix}.
\label{eq:SM_Kred_final}
\end{equation}
The full reduced Gaussian master equation for the two-mode system is therefore
\begin{equation}
\dot\rho_{ab}
=
-i[\hat H_{\rm red}(t),\rho_{ab}]
+
\sum_{i,j=a,b}
(\mathsf K_{\rm red})_{ij}
\left(
\hat d_i \rho_{ab} \hat d_j^\dagger
-\frac12\{\hat d_j^\dagger \hat d_i,\rho_{ab}\}
\right),
\label{eq:SM_ME_reduced_final}
\end{equation}
where
\begin{equation}
\hat{\mathbf d}
=
\begin{pmatrix}
\hat a\\
\hat b
\end{pmatrix},
\qquad
\hat H_{\rm red}(t)
=
\delta_a\,\hat a^\dagger \hat a
+\delta_b\,\hat b^\dagger \hat b
+
\mathcal E\,\hat a^{\dagger 2}
+
\mathcal E^\ast \hat a^2
+
\varepsilon(t)(\hat a+\hat a^\dagger).
\label{eq:SM_Hred_final}
\end{equation}
From Eq.~\eqref{eq:SM_ME_reduced_final}, the first moments satisfy
\begin{equation}
\begin{aligned}
\frac{d}{dt}\langle\hat{a}\rangle&=-(\gamma_{a}+i\delta_{a})\langle\hat{a}\rangle+g\,\langle\hat{b}\rangle-2i\mathcal{E}\,\langle\hat{a}^{\dagger}\rangle-i\varepsilon(t),&\frac{d}{dt}\langle\hat{a}^{\dagger}\rangle&=-(\gamma_{a}-i\delta_{a})\langle\hat{a}^{\dagger}\rangle+g^{\ast}\langle\hat{b}^{\dagger}\rangle+2i\mathcal{E}^{\ast}\langle\hat{a}\rangle+i\varepsilon(t),\\\frac{d}{dt}\langle\hat{b}\rangle&=-(\gamma_{b}+i\delta_{b})\langle\hat{b}\rangle+g^{\ast}\langle\hat{a}\rangle,&\frac{d}{dt}\langle\hat{b}^{\dagger}\rangle&=-(\gamma_{b}-i\delta_{b})\langle\hat{b}^{\dagger}\rangle+g\langle\hat{a}^{\dagger}\rangle.
\end{aligned}
\end{equation}
Introducing the normally ordered second moments
\begin{equation}
n_a=\langle \hat a^\dagger \hat a\rangle,
\qquad
n_b=\langle \hat b^\dagger \hat b\rangle,
\qquad
s_{ab}=\langle \hat a^\dagger \hat b\rangle,
\qquad
u_a=\langle \hat a^2\rangle,
\qquad
u_b=\langle \hat b^2\rangle,
\qquad
u_{ab}=\langle \hat a \hat b\rangle,
\end{equation}
the equations of motion are
\begin{equation}
\begin{aligned}
\dot{n}_{a}&=-2\gamma_{a}n_{a}+g\,s_{ab}+g^{\ast}s^{\ast}_{ab}-2i\mathcal{E}\,u^{\ast}_{a}+2i\mathcal{E}^{\ast}u_{a}-i\varepsilon(t)\langle\hat{a}^{\dagger}\rangle+i\varepsilon(t)\langle\hat{a}\rangle,\\\dot{n}_{b}&=-2\gamma_{b}n_{b}+g^{\ast}s_{ab}+g\,s^{\ast}_{ab},\\\dot{s}_{ab}&=\bigl[-(\gamma_{a}+\gamma_{b})+i(\delta_{a}-\delta_{b})\bigr]s_{ab}+g^{\ast}(n_{a}+n_{b})+2i\mathcal{E}^{\ast}u_{ab}+i\varepsilon(t)\langle\hat{b}\rangle,\\\dot{u}_{a}&=-2(\gamma_{a}+i\delta_{a})\,u_{a}+2g\,u_{ab}-2i\mathcal{E}(2n_{a}+1)-2i\varepsilon(t)\langle\hat{a}\rangle,\\\dot{u}_{b}&=-2(\gamma_{b}+i\delta_{b})\,u_{b}+2g^{\ast}u_{ab},\\\dot{u}_{ab}&=\bigl[-(\gamma_{a}+\gamma_{b})-i(\delta_{a}+\delta_{b})\bigr]u_{ab}+g\,u_{b}+g^{\ast}u_{a}-2i\mathcal{E}\,s_{ab}-i\varepsilon(t)\langle\hat{b}\rangle.
\end{aligned}
\end{equation}
The conjugate equations follow from complex conjugation.
These equations are exact consequences of the reduced Gaussian generator \eqref{eq:SM_ME_reduced_final}, and the closure follows from the Gaussian structure of the reduced generator.



\subsection{General drift matrix and rank-one pump structure}

Let \(m(t)\) denote the vector of annihilation-operator first moments. For the two-mode models,
\begin{equation}
m^{(2)}(t)=
\begin{pmatrix}
\langle \hat a\rangle\\
\langle \hat b\rangle
\end{pmatrix},
\end{equation}
whereas for the three-mode model,
\begin{equation}
m^{(3)}(t)=
\begin{pmatrix}
\langle \hat a\rangle\\
\langle \hat b\rangle\\
\langle \hat c\rangle
\end{pmatrix}.
\end{equation}
Because the pump Hamiltonian \eqref{eq:Hpump_general} is quadratic, the first moments do not close on \(m^{(k)}(t)\) alone; they couple to the conjugate first moments \(m^{(k)\ast}(t)\). It is therefore convenient to introduce the doubled vector
\begin{equation}
\mathbf v^{(k)}(t)=
\begin{pmatrix}
m^{(k)}(t)\\
m^{(k)\ast}(t)
\end{pmatrix}.
\end{equation}
For quadratic Hamiltonians and linear Lindblad jump operators, the homogeneous first-moment dynamics takes the form
\begin{equation}
\dot{\mathbf v}^{(k)}(t)=\mathcal A^{(k)}(\mathcal E)\,\mathbf v^{(k)}(t),
\qquad
\mathcal A^{(k)}(\mathcal E)=
\begin{pmatrix}
M^{(k)} & Q(\mathcal E)\\
Q(\mathcal E)^\ast & M^{(k)\ast}
\end{pmatrix},
\label{eq:bog_drift_general}
\end{equation}
where \(M^{(k)}\) is the annihilation-sector drift matrix in the absence of the two-photon pump, and \(Q(\mathcal E)\) contains the pairing terms generated by the pump.

For the single-mode two-photon pump \eqref{eq:Hpump_general} acting on mode \(\hat a\), the pump contribution to the first moments is given
\begin{equation}
\frac{d}{dt}\langle \hat a\rangle_{\rm pump}=-2i\mathcal E\,\langle \hat a^\dagger\rangle,
\qquad
\frac{d}{dt}\langle \hat a^\dagger\rangle_{\rm pump}=2i\mathcal E^\ast\,\langle \hat a\rangle,
\end{equation}
with no direct pairing contribution on \(\hat b\) or \(\hat c\). Therefore \(Q(\mathcal E)\) has rank one and, in the natural mode basis, takes the form
\begin{equation}
Q(\mathcal E)=\mathrm{diag}(-2i\mathcal E,0,\ldots,0).
\label{eq:Q_rank1}
\end{equation}

The broken regime is defined by the existence of an eigenvalue of \(\mathcal A^{(k)}(\mathcal E)\) with a positive real part:
\begin{equation}
\max_j \Re[\lambda_j(\mathcal A^{(k)}(\mathcal E))]>0.
\label{eq:broken_condition_general}
\end{equation}
Therefore, at the stability boundary ($\Re[\lambda]=0$), at least one eigenvalue lies on the imaginary axis, so we write
\begin{equation}
\lambda=i\omega,
\qquad
\omega\in\mathbb R,
\label{eq:s_iw}
\end{equation}
where \(\omega\) is the oscillation frequency of the first mode that reaches the boundary. In particular, when \(\omega=0\), the critical eigenvalue reaches \(\lambda=0\), and the instability is non-oscillatory. 
This indicates that, at this stationary solution, perturbations along the critical direction no longer decay.
%


\subsubsection{Exact determinant identity and general threshold formula}

We are interested in finding the following characteristic equation: $\det\!\left[\mathcal{A}^{(k)}(\mathcal{E})-\lambda\mathbb{I}\right]=0$. First, let us define
\begin{equation}
X_k(\lambda):=M^{(k)}-\lambda\mathbb I.
\end{equation}
Using the Schur complement for the block matrix \(\mathcal A^{(k)}(\mathcal E)-\lambda\mathbb I\), one finds
\begin{equation}
\det\!\left[\mathcal{A}^{(k)}(\mathcal{E})-\lambda\mathbb{I}\right]=\det\!\left[X_k(\lambda)\right]\det\!\left[X_k^{*}(\lambda^{*})-Q^{\ast}(\mathcal{E})X_k^{-1}(\lambda)Q(\mathcal{E})\right].
\label{eq:schur_general}
\end{equation}
Since \(Q(\mathcal E)\) has rank one and only its \((1,1)\) entry is nonzero, the matrix determinant lemma yields
\begin{equation}
\det\!\left[\mathcal{A}^{(k)}(\mathcal{E})-\lambda\mathbb{I}\right]=\det\!\left[X_k(\lambda)\right]\det\!\left[X_k^{*}(\lambda^{*})\right]-4|\mathcal{E}|^{2}C_{11}\!\left[X_k(\lambda)\right]C_{11}\!\left[X_k^{*}(\lambda^{*})\right],
\label{eq:det_identity_general}
\end{equation}
where
\begin{equation}
C_{11}\left[X_k\right]=\bigl[\operatorname{adj}(X_k)\bigr]_{11}
\end{equation}
is the \((1,1)\) cofactor of \(X\), namely the determinant of the minor obtained by deleting the first row and first column, with the standard sign convention. Here, \(\operatorname{adj}(X)\) denotes the adjugate matrix, satisfying
\begin{equation}
\operatorname{adj}(X_k)=\det\left[X_k\right]\,X_k^{-1}
\end{equation}
whenever \(X_k\) is invertible.
Finally, it is convenient to define
\begin{equation}
p_k(\lambda):=\det\!\left[X_k(\lambda)\right],\qquad q_k(\lambda):=C_{11}\!\left[X_k(\lambda)\right].
\label{eq:pq_general_defs}
\end{equation}
Thus, since
\begin{equation}
\det\!\left[X_k^{*}(\lambda^{*})\right]=p_k^{\ast}(\lambda^{\ast}),\qquad C_{11}\!\left[X_k^{*}(\lambda^{*})\right]=q_k^{\ast}(\lambda^{\ast}),
\label{eq:conj_general_rel}
\end{equation}
Eq.~\eqref{eq:det_identity_general} becomes
\begin{equation}
\det\!\left[\mathcal{A}^{(k)}(\mathcal{E})-\lambda\mathbb{I}\right]
=
p_k(\lambda)p_k^\ast(\lambda^\ast)
-
4|\mathcal E|^2\,q_k(\lambda)q_k^\ast(\lambda^\ast).
\label{eq:det_identity_pq_general}
\end{equation}
Therefore, the characteristic equation is
\begin{equation}
p_k(\lambda)p_k^\ast(\lambda^\ast)
=
4|\mathcal E|^2\,q_k(\lambda)q_k^\ast(\lambda^\ast),
\label{eq:char_eq_general}
\end{equation}
or equivalently,
\begin{equation}
|\mathcal E|^2
=
\frac{p_k(\lambda)p_k^\ast(\lambda^\ast)}
{4\,q_k(\lambda)q_k^\ast(\lambda^\ast)}.
\label{eq:E_general_from_lambda}
\end{equation}
At the instability threshold we set \(\lambda=i\omega\), \(\omega\in\mathbb R\). For each fixed \(\omega\), Eq.~\eqref{eq:E_general_from_lambda} gives the pump amplitude required for an eigenvalue to sit exactly on the imaginary axis at that frequency:
\begin{equation}
\mathcal E^{(k)}_{\rm crit}(\omega)
:=
\frac12
\sqrt{
\frac{p_k(i\omega)\,p_k^\ast(-i\omega)}
{q_k(i\omega)\,q_k^\ast(-i\omega)}
}.
\label{eq:Ecrit_general_omega}
\end{equation}
For fixed system parameters, the physical threshold is the smallest value of \(|\mathcal E|\) for which the boundary condition is satisfied for at least one real \(\omega\). Hence
\begin{equation}
\mathcal E^{(k)}_{\rm crit}
=
\min_{\omega\in\mathbb R}\mathcal E^{(k)}_{\rm crit}(\omega).
\label{eq:Ecrit_general_min}
\end{equation}
No further simplification is assumed at this stage. Any reduction of Eq.~\eqref{eq:Ecrit_general_omega} to a modulus-squared form must be checked directly from the explicit model-dependent expressions for \(p\) and \(q\).

\subsubsection{Case I: Two-mode coherent interaction (\(\Gamma=0\), \(J\neq 0\))}
\label{subsec:case_coherent}

Here, we consider two modes \(\hat a\) and $\hat b$, with Hamiltonians
\begin{equation}
\hat{H}_{0}=\delta_{a}\,\hat{a}^{\dagger}\hat{a}+\delta_{b}\,\hat{b}^{\dagger}\hat{b},\qquad\hat H_{\rm int}^{\rm coh}=J\,\hat{a}^{\dagger}\hat{b}+J^{\ast}\hat{a}\hat{b}^{\dagger},\qquad\hat H_{\rm pump}=\mathcal E\,\hat a^{\dagger 2}+\mathcal E^\ast \hat a^2,
\end{equation}
together with local Lindblad losses \(\kappa_a \mathcal D[\hat a]+\kappa_b\mathcal D[\hat b]\), and no shared dissipative reservoir. The annihilation-sector drift matrix \(M^{(2)}\) acting on
\begin{equation}
m^{(2)}=
(\langle\hat{a}\rangle,\langle\hat{b}\rangle)^{T}
\end{equation}
is
\begin{equation}
M^{(2)}=
\begin{pmatrix}
-\gamma_a-i\delta_a & -iJ\\
-iJ^\ast & -\gamma_b-i\delta_b
\end{pmatrix},
\label{eq:M2_coherent}
\end{equation}
where $\gamma_{a,b}=\kappa_{a,b}/2$. Hence, one can show that
\begin{equation}
    p_{2}(\lambda)=\det(M^{(2)}-\lambda\mathbb{I})=(\gamma_{a}+i\delta_{a}+\lambda)(\gamma_{b}+i\delta_{b}+\lambda)+|J|^{2}
\end{equation}
and
\begin{equation}
    q_{2}(\lambda)=C_{11}(M^{(2)}-\lambda\mathbb{I})=[\textrm{adj}(M^{(2)}-\lambda\mathbb{I})]_{11}=-\gamma_{b}-i\delta_{b}-\lambda.
\end{equation}
Now, from Eq.~\eqref{eq:Ecrit_general_omega}, we obtain
\begin{equation}
\mathcal{E}^{(2)}_{{\rm crit}}(\omega)=\frac{1}{2}\sqrt{\frac{\left[\left(\gamma_{a}+i\left(\delta_{a}+\omega\right)\right)\left(\gamma_{b}+i\left(\delta_{b}+\omega\right)\right)+|J|^{2}\right]\,\left[\left(\gamma_{a}-i\left(\delta_{a}-\omega\right)\right)\left(\gamma_{b}-i\left(\delta_{b}-\omega\right)\right)+|J|^{2}\right]}{\delta^{2}_{b}+(\gamma_{b}+i\omega)^{2}}}.
\label{eq:Ecrit_coh_explicit_general}
\end{equation}


\subsubsection{Case II: Three-mode dissipative engineered interaction (\(\Gamma\neq0\), \(J= 0\))}
\label{subsec:case_3mode_diss}

We now consider the scenario of three modes (\(\hat a,\hat b,\hat c\)) with two-photon pump in mode \(\hat a\), dissipative engineered interactions mediated by \(\hat c\) and no coherent coupling between \(\hat a\) and \(\hat b\) ($J=0$). Hence
\begin{equation}
    \hat{H}_{3{\rm m}}=\delta_{a}\hat{a}^{\dagger}\hat{a}+\delta_{b}\hat{b}^{\dagger}\hat{b}+\delta_{c}\hat{c}^{\dagger}\hat{c}+\mathcal{E}\hat{a}^{\dagger2}+\mathcal{E}^{\ast}\hat{a}^{2}.
\end{equation}
For the common reservoir, the jump operators are given by
\begin{equation}\label{zab}
\hat z_a=p_a \hat a+p_{ca}\hat c,
\qquad
\hat z_b=p_b \hat b+p_{cb}\hat c,
\end{equation}
supplemented by local losses in all three modes. 
Under these conditions, the annihilation-sector drift matrix acting on $m^{(3)}=(\langle\hat{a}\rangle,\langle\hat{b}\rangle,\langle\hat{c}\rangle)^{T}$ is written as
\begin{equation}
M^{(3)}=\begin{pmatrix}-\gamma_{a}-i\delta_{a} & 0 & -g_{ac}\\
0 & -\gamma_{b}-i\delta_{b} & -g_{bc}\\
-g^{*}_{ac} & -g^{*}_{bc} & -\gamma_{c}-i\delta_{c}
\end{pmatrix},
\label{eq:M3_diss_topology}
\end{equation}
where $\gamma_{a,b}=\left(\kappa_{a,b}+\Gamma|p_{a,b}|^{2}\right)/2$, $\gamma_{c}=\left(\kappa_{c}+\Gamma|p_{ca}|^{2}+\Gamma|p_{cb}|^{2}\right)/2$ are the effective dissipative couplings between \(a\) and \(c\), and between \(b\) and \(c\), respectively. Here, $g_{a(b)c}=\Gamma p^{*}_{a(b)}p_{ca(b)}/2$ denotes the off-diagonal coefficient of the first-moment drift matrix generated by the common dissipators. Note that \(g_{ab}=0\) since there is no jump operator containing both \(\hat a\) and \(\hat b\).
Given Eq.~\eqref{eq:M3_diss_topology}, one can compute the \((1,1)\) cofactor
\begin{equation}\label{eq:q3_lambda_raw}
    q_{3}(\lambda)=C_{11}(M^{(3)}-\lambda\mathbb{I})=[\textrm{adj}(M^{(3)}-\lambda\mathbb{I})]_{11}=(\gamma_{b}+i\delta_{b}+\lambda)(\gamma_{c}+i\delta_{c}+\lambda)-|g_{bc}|^{2},
\end{equation}
and
\begin{equation}\label{eq:p3_lambda_raw}
    p_{3}(\lambda)=\det(M^{(3)}-\lambda\mathbb{I})=-(\gamma_{a}+i\delta_{a}+\lambda)q_{3}(\lambda)+(\gamma_{b}+i\delta_{b}+\lambda)|g_{ac}|^{2}.
\end{equation}
Substituting Eqs.~\eqref{eq:q3_lambda_raw} and \eqref{eq:p3_lambda_raw} into the general threshold formula \eqref{eq:Ecrit_general_omega} gives
\begin{equation}
\mathcal E_{\rm crit}^{(3)}(\omega) = \frac12\sqrt{\frac{
p_3(i\omega)\,p_3^\ast(-i\omega)}{q_3(i\omega)\,q_3^\ast(-i\omega)}},
\qquad
\mathcal E_{\rm crit}^{(3)} = \min_{\omega\in\mathbb R}\mathcal E_{\rm crit}^{(3)}(\omega),
\label{eq:Ecrit_3mode_generalform}
\end{equation}
where
\begin{equation}
\begin{aligned}
    q_{3}(i\omega)&=(\gamma_{b}+i(\delta_{b}+\omega))(\gamma_{c}+i(\delta_{c}+\omega))-|g_{bc}|^{2},\\
    q^{*}_{3}(-i\omega)&=(\gamma_{b}-i(\delta_{b}-\omega))(\gamma_{c}-i(\delta_{c}-\omega))-|g_{bc}|^{2},\\
    p_{3}(i\omega)&=-(\gamma_{a}+i(\delta_{a}+\omega))q_{3}(i\omega)+(\gamma_{b}+i(\delta_{b}+\omega))|g_{ac}|^{2},\\
    p^{*}_{3}(-i\omega)&=-(\gamma_{a}-i(\delta_{a}-\omega))q^{*}_{3}(-i\omega)+(\gamma_{b}-i(\delta_{b}-\omega))|g_{ac}|^{2}.
\end{aligned}
\end{equation}
%
%
%


\subsubsection{Case III: Reduced two-mode dissipative model (\(\Gamma\neq0\), \(J= 0\) and mediator eliminated)}
\label{subsec:case_reduced_diss}

Adiabatically eliminating the auxiliary mode \(\hat c\) yields the following effective reduced two-mode dissipative drift matrix in the annihilation sector,
\begin{equation}\label{eq:Mred_def}
M^{(2)}_{\textrm{red}}=\begin{pmatrix}-\gamma_{a}-i\delta_{a} & g\\
g^{*} & -\gamma_{b}-i\delta_{b}
\end{pmatrix},\qquad g \coloneqq\frac{\Gamma_{\textrm{eff}}\mu_{ca}\mu^{*}_{cb}}{2},\qquad\gamma_{j} \coloneqq\frac{\kappa_{j}}{2}+\frac{\Gamma_{j}}{2}-\frac{|\mu_{cj}|^{2}}{2}\Gamma_{\textrm{eff}},
\end{equation}
with
\begin{equation}
    \mu_{ck} \coloneqq p_{ck}p^{*}_{k},\qquad\Gamma_{k}\coloneqq \Gamma|p_{k}|^{2},\qquad\Gamma_{ck} \coloneqq\Gamma|p_{ck}|^{2},\qquad \Gamma_{\textrm{eff}}\coloneqq\frac{\Gamma^{2}}{\left(\kappa_{c}+\Gamma_{ca}+\Gamma_{cb}\right)}.
\end{equation}

Therefore, 
\begin{equation}
    p_{\textrm{red}}(\lambda)=\det(M^{(2)}_{\textrm{red}}-\lambda\mathbb{I})=(\gamma_{a}+i\delta_{a}+\lambda)(\gamma_{b}+i\delta_{b}+\lambda)-|g|^{2},
\end{equation}
and
\begin{equation}
    q_{\textrm{red}}(\lambda)=C_{11}(M^{(2)}_{\textrm{red}}-\lambda\mathbb{I})=[\textrm{adj}(M^{(2)}_{\textrm{red}}-\lambda\mathbb{I})]_{11}=-\gamma_{b}-i\delta_{b}-\lambda,
\end{equation}
with
\begin{equation}
\begin{aligned}
p_{\textrm{red}}(i\omega)&=(\gamma_{a}+i(\delta_{a}+\omega))(\gamma_{b}+i(\delta_{b}+\omega))-|g|^{2},\qquad&q_{\textrm{red}}(i\omega)&=-\gamma_{b}-i(\delta_{b}+\omega),\\p^{*}_{\textrm{red}}(-i\omega)&=(\gamma_{a}-i(\delta_{a}-\omega))(\gamma_{b}-i(\delta_{b}-\omega))-|g|^{2},\qquad&q^{*}_{\textrm{red}}(-i\omega)&=-\gamma_{b}+i(\delta_{b}-\omega).
\end{aligned}
\end{equation}
Finally, the exact threshold curve reads
\begin{equation}
\mathcal E_{\rm crit}^{({\rm red})}(\omega)
=
\frac12
\sqrt{
\frac{
p_{\rm red}(i\omega)\,p_{\rm red}^\ast(-i\omega)
}{
q_{\rm red}(i\omega)\,q_{\rm red}^\ast(-i\omega)
}
},
\qquad
\mathcal E_{\rm crit}^{({\rm red})}
=
\min_{\omega\in\mathbb R}\mathcal E_{\rm crit}^{({\rm red})}(\omega).
\label{eq:Ecrit_reduced_generalform}
\end{equation}


\subsection{Passive limit and absence of autonomous growth}

It is important to distinguish the passive dissipative sector from the pump-induced instability. When the two-photon pump is absent, $\mathcal E=0$, the reduced first-moment drift matrix is
\begin{equation}
\dot m^{(2)}(t)=M_{\rm red}^{(2)}m^{(2)}(t).
\end{equation}
The corresponding reduced Kossakowski matrix is
\begin{equation}
\mathsf K_{\rm red}
=
\begin{pmatrix}
2\gamma_a & -2g\\
-2g^\ast & 2\gamma_b
\end{pmatrix}.
\end{equation}
Because $\mathsf K_{\rm red}$ is the Schur complement of the positive three-mode Kossakowski matrix, one has
\begin{equation}
\mathsf K_{\rm red}\ge 0.
\end{equation}
Equivalently,
\begin{equation}
\gamma_a\ge 0,
\qquad
\gamma_b\ge 0,
\qquad
\gamma_a\gamma_b\ge |g|^2 .
\label{eq:passive_constraint}
\end{equation}
Moreover,
\begin{equation}
\frac{M_{\rm red}^{(2)}+M_{\rm red}^{(2)\dagger}}{2}
=
-
\frac{\mathsf K_{\rm red}}{2}
\le 0.
\label{eq:Hermitian_part_passive}
\end{equation}
Therefore all eigenvalues of $M_{\rm red}^{(2)}$ have non-positive real parts. If $\mathsf K_{\rm red}$ is positive definite, the real parts are strictly negative and all first moments decay. Thus, the passive reduced model cannot display autonomous exponential growth when $\mathcal E=0$, even if the initial conditions $a(0)$ or $b(0)$ are nonzero.

This is the precise sense in which the reservoir-engineered dissipative sector remains passive. The broken regime discussed in the main text appears only after the two-photon pump is included, because the pump adds the pairing block $Q(\mathcal E)$ to the drift,
\begin{equation}
\mathcal A_{\rm red}^{(2)}(\mathcal E)
=
\begin{pmatrix}
M_{\rm red}^{(2)} & Q(\mathcal E)\\
Q^\ast(\mathcal E) & M_{\rm red}^{(2)\ast}
\end{pmatrix}.
\end{equation}
The instability threshold is therefore a pump-induced stability threshold of the physical drift matrix. The energy source for exponential growth is the external two-photon pump, while the passive reservoir-engineered interaction lowers the pump amplitude required to reach this threshold.

\begin{figure}[t]
    \centering
    \includegraphics[width=1\columnwidth]{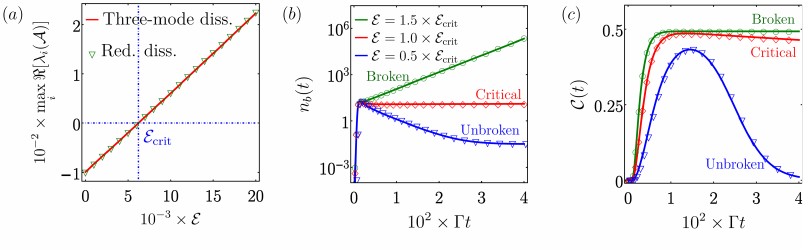}
    \caption{Comparison between the three-mode dissipative engineered topology and its respective reduced two-mode dissipative model. (a) $\max_i\Re[\lambda_i(\mathcal A(\mathcal E))]$ as a function of the two-photon drive amplitude \(\mathcal E\). (b) Charging dynamics on a logarithmic scale, and (c) coherent fraction $\mathcal{C}(t)\coloneq |\langle \hat b(t)\rangle|^2/n_b(t)$ below, at, and above the parametric-instability threshold \(\mathcal E_{\rm crit}\) for the three-mode (continuous) and reduced (markers) models. The reduced model matches the three-mode reference.
    Parameters used: $\delta_a=\delta_b=\delta_c=0$, \(\kappa_a=\kappa_b=0.02\), $\kappa_c=0$, \(\Gamma=0.4\), \(p_a=1\), \(p_b=2\), \(p_{ca}=p_{cb}=10\), \(\varepsilon_0=1\), \(t_0=20\), and \(\sigma=5\).\justifying}
    \label{Comparison}
\end{figure}

\subsection{Comparison protocol and growth-rate curves}
\label{subsec:comparison_protocol}
For each scenario, define the pump-dependent drift matrix \(\mathcal A(\mathcal E)\) via Eq. \eqref{eq:bog_drift_general} with the appropriate \(M^{(k)}\) and with \(Q(\mathcal E)\) given by Eq. \eqref{eq:Q_rank1}. The growth-rate function $\max_i \Re[\lambda_i(\mathcal A(\mathcal E))]$ determines stability. The pump threshold \(\mathcal E_{\rm crit}\) is equivalently characterized as (i) the one-dimensional minimization \eqref{eq:Ecrit_general_min} (with model-specific \(p,q\)), or (ii) the zero-crossing of \(\max_i \Re[\lambda_i(\mathcal A(\mathcal E))]\). 
In Fig. \ref{Comparison}, we plot \(\max_i \Re[\lambda_i(\mathcal A(\mathcal E))]\) versus \(\mathcal E\) for the three models and mark the corresponding \(\mathcal E_{\rm crit}\) obtained from \eqref{eq:Ecrit_coh_explicit_general}, \eqref{eq:Ecrit_3mode_generalform}, and \eqref{eq:Ecrit_reduced_generalform}. 

For the parameter regime used in the main text, the auxiliary mode is chosen to be much faster than the charger and battery modes. The relevant hierarchy is then
\begin{equation}
\gamma_c
\gg
\left\{
\frac{\kappa_a + \Gamma_a}{2},\;
\frac{\kappa_b + \Gamma_b}{2},\;
|g_{ac}|,\;
|g_{bc}|,\;
|\delta_{a(b)}|,\;
|\mathcal E|,\;
\sigma^{-1},\;
\max_i \Re[\lambda_i(\mathcal A(\mathcal E))]
\right\}.
\end{equation}

\suppnote{Pump-power economy and coherent-work yield}
\label{sec:SM_pump_power_economy}

Here, we provide the analytical definitions behind the pump-power discussion in the main text. We use \(\nu\in\{{\rm red},{\rm coh}\}\) to label the reduced dissipative architecture and the coherent beam-splitter benchmark. The pump-dependent drift matrix of architecture \(\nu\) is denoted by \(\mathcal A_\nu(\mathcal E)\), and its respective spectral abscissa reads
\begin{equation}
\alpha_\nu(\mathcal E)
=
\max_j{\Re}[\lambda_j(\mathcal A_\nu(\mathcal E))] .
\label{eq:SM_spectral_abscissa_power}
\end{equation}
The instability threshold is defined by the smallest pump amplitude for which the spectral abscissa becomes positive, i.e.,
\begin{equation}
\mathcal E_{\rm crit}^{\nu}
=
\inf
\left\{
|\mathcal E|:
\alpha_\nu(\mathcal E)>0
\right\}.
\label{eq:SM_threshold_definition_power}
\end{equation}
In the numerical comparisons, the coherent benchmark is calibrated at equal effective charger--battery coupling strength, \(|J|\equiv|g|\), so that differences in the threshold originate from the structure of the drift matrix rather than from a larger interaction scale.

\subsection{From pump amplitude to pump power and pump-power operation window}

The two-photon pump Hamiltonian is given by
\begin{equation}
\hat H_{\rm pump}
=
\mathcal E\hat a^{\dagger 2}
+
\mathcal E^\ast\hat a^2 .
\label{eq:SM_H_pump_power}
\end{equation}
For amplitude-linear parametric implementations, such as flux-modulated superconducting Kerr resonators, the two-photon drive coefficient is proportional to the applied classical pump amplitude. Therefore, the physical pump power is proportional to \(|\mathcal E|^2\), up to a platform-dependent calibration factor \(\chi_{\rm p}\) \cite{Beaulieu2025NatCommun,Krantz2016NatCommun}, i.e.,
\begin{equation}
P_{\rm pump}
=
\chi_{\rm p}|\mathcal E|^2 ,
\label{eq:SM_pump_power_calibration}
\end{equation}
The same \(\chi_{\rm p}\) is assumed for the reduced dissipative architecture and for the coherent benchmark when the two models are compared using the same pump convention. 
From Eq.~\eqref{eq:SM_pump_power_calibration} one can define the platform-independent critical pump-power ratio as
\begin{equation}
\mathcal R_P
=
\frac{P_{\rm crit}^{\rm red}}{P_{\rm crit}^{\rm coh}}
=
\left|
\frac{\mathcal E_{\rm crit}^{\rm red}}
{\mathcal E_{\rm crit}^{\rm coh}}
\right|^2 ,
\label{eq:SM_power_ratio}
\end{equation}
The corresponding critical-power saving is $ \mathcal S_P
= 1-\mathcal R_P $. The same saving can be expressed in decibels as
\begin{equation}
\Delta_{\rm dB}
=
10\log_{10}
\left(
\frac{P_{\rm crit}^{\rm coh}}
{P_{\rm crit}^{\rm red}}
\right)
=
20\log_{10}
\left|
\frac{\mathcal E_{\rm crit}^{\rm coh}}
{\mathcal E_{\rm crit}^{\rm red}}
\right| .
\label{eq:SM_db_saving}
\end{equation}
For the representative parameter set used in Fig.~\ref{Fig2} of the main text,
\(
\mathcal E_{\rm crit}^{\rm red}
\approx
6.2\times 10^{-3},
\)
\(
\mathcal E_{\rm crit}^{\rm coh}
\approx
1.0\times 10^{-2},
\)
Eqs.~\eqref{eq:SM_power_ratio}--\eqref{eq:SM_db_saving} give
\(
\mathcal R_P
\simeq
0.38,
\)
\(
\mathcal S_P
\simeq
0.61,
\)
and
\(
\Delta_{\rm dB}
\simeq
4.1\,{\rm dB}.
\)
Thus, the reduced dissipative architecture reaches the broken regime with about \(61\%\) less critical pump power than the coherent benchmark with \(|J|\equiv|g|\).

We can interpret the corresponding critical power saving $\mathcal{S}_P$ in another way. To see this, consider the ordering \(\mathcal E_{\rm crit}^{\rm red}<\mathcal E_{\rm crit}^{\rm coh}\) that defines an interval of pump amplitudes in which the two architectures are in different dynamical regimes:
\begin{equation}
\mathcal E_{\rm crit}^{\rm red}
<
|\mathcal E|
<
\mathcal E_{\rm crit}^{\rm coh}.
\label{eq:SM_amplitude_window}
\end{equation}
Using Eq.~\eqref{eq:SM_pump_power_calibration}, the same window can be written in terms of the pump power as
\(
P_{\rm crit}^{\rm red}
<
P_{\rm pump}
<
P_{\rm crit}^{\rm coh}
\). Accordingly, the absolute width of the window reads
\begin{equation}
\Delta P_{\rm win}
=
P_{\rm crit}^{\rm coh}
-
P_{\rm crit}^{\rm red}
=
\chi_{\rm p}
\left(
|\mathcal E_{\rm crit}^{\rm coh}|^2
-
|\mathcal E_{\rm crit}^{\rm red}|^2
\right).
\label{eq:SM_power_window_width}
\end{equation}
The normalized width relative to the coherent benchmark threshold is
\begin{equation}
\frac{\Delta P_{\rm win}}{P_{\rm crit}^{\rm coh}}
= 1-\mathcal R_P
= \mathcal S_P.
\label{eq:SM_normalized_window_width}
\end{equation}
This implies that \(\mathcal S_P\) has two distinct but equivalent meanings: it is the fractional reduction of the critical pump power, and it also quantifies the fraction of the coherent threshold power over which the reduced dissipative architecture is already in the broken regime while the coherent benchmark remains below the threshold.

\end{document}